\documentclass[12pt]{article}
\usepackage{latexsym,epsfig,graphicx,epstopdf,amsmath,amssymb,amscd,undertilde,multirow,chicago,psfrag,paralist,dsfont,url, listings, booktabs, algorithm, algorithmicx, algpseudocode, amssymb, amsmath, bm}

\usepackage[skip=5pt]{caption}
\usepackage[titletoc]{appendix}
\usepackage[T1]{fontenc}
\usepackage[american]{babel}
\usepackage{comment}
\usepackage{varwidth}
\usepackage{verbatim}

\newcommand{\veps}{\varepsilon}

\newcommand{\NOR}{{\rm N}}

\newcommand{\betahat}{\widehat{\beta}}

\newcommand{\xvec}{\boldsymbol{x}}

\newcommand{\xbar}{\overline{x}}

\newcommand{\D}{\mathcal{D}}
\newcommand{\minitab}[2][l]{\begin{tabular}{#1}#2\end{tabular}}

\def\baselinestretch{1.25}

\begin{document}

\title{Performance Evaluation of Large Language Models in Statistical Programming}

\author{
Xinyi Song$^{1}$, Kexin Xie$^{1}$, Lina Lee$^{1}$,  Ruizhe Chen$^{1}$, Jared M. Clark$^{1}$,\\ Hao He$^{1}$, Haoran He$^{1}$, Jie Min$^{2}$, Xinlei Zhang$^{1}$, Simin Zheng$^{1}$,\\ Zhiyang Zhang$^{1}$, Xinwei Deng$^{1}$, and Yili Hong$^{1}$\\[1.5ex]
{\small $^{1}$Department of Statistics, Virginia Tech, Blacksburg, VA 24061}\\
{\small $^{2}$Department of Mathematics \& Statistics, University of South Florida, Tampa, FL 33620}
}

\date{}

\maketitle
\begin{abstract}

The programming capabilities of large language models (LLMs) have revolutionized automatic code generation and opened new avenues for automatic statistical analysis. However, the validity and quality of these generated codes need to be systematically evaluated before they can be widely adopted. Despite their growing prominence, a comprehensive evaluation of statistical code generated by LLMs remains scarce in the literature. In this paper, we assess the performance of LLMs, including two versions of ChatGPT and one version of Llama, in the domain of SAS programming for statistical analysis. Our study utilizes a set of statistical analysis tasks encompassing diverse statistical topics and datasets. Each task includes a problem description, dataset information, and human-verified SAS code. We conduct a comprehensive assessment of the quality of SAS code generated by LLMs through human expert evaluation based on correctness, effectiveness, readability, executability, and the accuracy of output results. The analysis of rating scores reveals that while LLMs demonstrate usefulness in generating syntactically correct code, they struggle with tasks requiring deep domain understanding and may produce redundant or incorrect results. This study offers valuable insights into the capabilities and limitations of LLMs in statistical programming, providing guidance for future advancements in AI-assisted coding systems for statistical analysis.

\end{abstract}

\newpage

\section{Introduction}\label{sec:introudction}
\subsection{Background and Motivation}
Artificial intelligence (AI) has become increasingly important in many areas, and its growth continues at an unprecedented pace. In recent years, the rise of large language models (LLMs) has led to significant advancements in AI-generated coding tools such as ChatGPT (\shortciteNP{openai2024gpt4technicalreport}) and Llama (\shortciteNP{touvron2023llamaopenefficientfoundation}). For instance, LLMs can generate statistical code in SAS (\citeNP{sassoftware}) and R (\citeNP{R}). These exciting developments offer the potential to reduce manual coding efforts, enhance accessibility to statistical analysis for non-statisticians, and potentially pave the way toward fully automated statistical analysis.

While LLMs can generate codes based on their generative models, the validity and quality of those generated code require thorough investigation, which is particularly important for non-expert users who lack expertise in statistical programming. Non-expert users may find it challenging to evaluate the correctness and reliability of AI-generated code and its outputs. Therefore, it is important to establish a systematic evaluation framework to investigate the quality of AI-generated code and ensure the results are reliable before using them in the workplace. This motivates us to evaluate the performance of LLMs in statistical programming by leverage the human evaluation from domain experts.

Human evaluation is the essential starting point for this effort. Although automatic metrics, such as those commonly used in natural language processing (NLP), can be employed to assess generated code, their applicability to statistical programming remains uncertain. Concerns persist about the reliability and trustworthiness of these metrics in this specific context. Consequently, human evaluation can serve as the gold standard for assessing the quality of generated statistical code.

Human evaluation of generated code is essential but comes with its challenges. The primary challenge is the significant demand for expert hours, as code can be complex and difficult to assess. We selected SAS as the starting point for human evaluation. SAS programming is a powerful tool for data management and analysis, widely used in industries such as finance and pharmaceutical industry. Its well-structured syntax facilitates evaluation. However, writing SAS code and interpreting its results require a deep understanding of programming syntax and statistical models, which can be challenging for those without substantial expertise. In this study, considerable human labor hours have been dedicated to expert evaluation of statistical code generated by LLMs, marking a pioneering effort in the literature. Specifically, more than 10 participants, consisting of statistics faculty and PhD candidates, have contributed to the evaluation, devoting hundreds of hours to the study.

The overarching goal of this paper is to evaluate the performance of LLMs in statistical programming, specifically in the context of SAS coding, through human expert assessments. We investigate the performance of three LLMs: GPT 3.5, GPT 4.0, and Llama 3.1 70B. To begin, we manually curated a set of 207 statistical analysis tasks, each consisting of a problem description, dataset details, and human-verified SAS code. These tasks were then input into the LLMs, which generated SAS code for each task. Human experts evaluated the generated code based on various criteria, including correctness, effectiveness, readability, executability, and the accuracy of the output results. Statistical analyses were conducted on the evaluation scores to address research questions regarding the performance of LLMs in statistical programming. The specific objectives of this study are as follows: 1) provide an overall summary of the performance of LLMs in statistical programming with SAS; 2) identify the strengths and weaknesses of LLMs; 3) compare the performance of different LLMs (i.e., ChatGPT versus Llama) and examine whether the version of the LLM makes a difference (i.e., ChatGPT 3.5 versus 4.0); and 4) explore potential areas for improvement in automatic statistical analysis.

\subsection{Related Literature and Contribution of This Work}

Generative AI has profoundly influenced the field of programming by enabling automated code generation and assisting with complex tasks. Early approaches to code generation leveraged sequence-to-sequence models to translate problem descriptions in natural language into executable code. \shortciteN{ling2016latentpredictornetworkscode} introduced latent predictor networks to map text to code, while \shortciteN{yin2017syntacticneuralmodelgeneralpurpose} proposed a syntactic neural network model that constructs abstract syntax trees, ensuring syntactic correctness. The transformer architecture, presented by \shortciteN{NIPS20173f5ee243}, revolutionized NLP by effectively capturing long-range dependencies. Building on this foundation, GPT 2.0 and GPT 3.0, pre-trained on extensive web corpora, achieved remarkable performance across diverse NLP tasks (\shortciteNP{brown2020languagemodelsfewshotlearners}), paving the way for the development of LLMs.

LLMs, such as ChatGPT, have the capability to automatically generate code in various programming languages using natural language prompts (\shortciteNP{mollick2022chatgpt}; \citeNP{a17020062}). A significant milestone in AI code generation was the development of OpenAI Codex, an LLM specifically trained on code datasets. Built upon Codex, GitHub Copilot serves as an AI-powered coding assistant, providing code suggestions for various tasks. Additionally, Meta AI introduced the Llama family of models, trained on extensive text datasets, followed by the release of Code Llama (\shortciteNP{bap2024codellamaopenfoundation}), which is fine-tuned on code-specific data to enhance its proficiency in code generation. These LLMs are widely used and significantly enhance programming efficiency.

To effectively deploy AI-generated code into production, it is crucial to use a systematic framework to evaluate its performance and reliability in terms of functional correctness, code quality, maintainability, and readability (\shortciteNP{hong2023statistical}; \shortciteNP{AIcode2024Rabbi}). \shortciteN{chen2021evaluatinglargelanguagemodels} found that NLP metrics like BLEU correlate poorly with functional correctness in code and recommended execution-based evaluations, introducing HumanEval as a benchmark with unit tests for assessing Codex's code quality. Similarly, \shortciteN{hendrycks2021measuringcodingchallengecompetence} assessed the problem-solving abilities of AI-generated code using the APPS dataset, focusing on correctness and efficiency through automated test cases. \shortciteN{vaith2022} evaluated the usability and effectiveness of tools like OpenAI Codex and GPT, revealing gaps between user expectations and actual performance. \shortciteN{Clark10608315} investigated the quality and consistency of AI-generated code, noting that while ChatGPT generally produces high-quality code, occasional bugs persist. Additionally, \shortciteN{chen2021evaluatinglargelanguagemodels} highlighted Codex's capability to handle complex Python tasks. Also, the multi-modal model offers a wide range of AI applications, presenting opportunities to further investigate its quality in AI-generated SAS code for statistical analysis (e.g., \shortciteNP{guo2023large}).

Regarding the performance of GPT 3.5, GPT 4.0, and Llama, \citeN{Atkingpt2023} analyzed GPT 3.5's capabilities across various programming languages, observing frequent inaccuracies in complex programming scenarios. Similarly, \citeN{sherman2024} evaluated GPT 3.5 using non-compliant C code examples, revealing that it frequently overlooked obvious errors and focused on minor issues while neglecting critical problems. \shortciteN{bubeck2023sparksartificialgeneralintelligence} found that GPT 4.0 outperforms GPT 3.5 in code generation tasks. Additionally, \shortciteN{comparison2024yeadon} compared human performance with GPT 3.5 and GPT 4.0 on university-level coding tasks, showing that GPT 4.0 excels over GPT 3.5. However, both models still fall short compared to human students, particularly in tasks requiring nuanced or context-specific understanding. \citeN{portakal2024llama3vsGPT4} assessed coding performance using the HumanEval benchmark, where the Llama 3 70B model achieved a score of 81.7\%, while GPT 4.0 outperformed it with a score of 85.9\%.

In data science and statistical programming, recent studies have integrated LLMs into analytics tasks and conducted performance evaluations. These studies primarily focus on specific aspects, such as particular model types (\shortciteNP{bordt2024datasciencellmsinterpretable}), data settings (\shortciteNP{Feng10196869}), software efficiency (\shortciteNP{NitinSherje2024}), and optimization (\shortciteNP{Ad10405148opt}). \shortciteN{Megahed02042024} studied how generative AI models (e.g., ChatGPT) can be used or misused in statistics practice. \shortciteN{Megahedetal2024SQC} introduced a chatbot system, called ChatSQC, that combines LLM with statistical quality control.  However, no systematic framework exists for evaluating the performance of LLMs in data science and statistical programming, nor have any studies examined the performance of LLM-generated code in statistical analysis.

The major contribution of this work is developing a systematic framework for evaluating the performance of LLMs in statistical programming. We establish a comprehensive dataset to assess LLM capabilities and employ human evaluations to generate performance ratings. Through comprehensive analysis of these evaluation scores, we provide insights into the strengths and weaknesses of LLMs, enabling benchmarking and identifying areas for improvement. Our findings also highlight areas for future improvement of LLMs.

\subsection{Overview}
The remainder of the paper is organized as follows. Section~\ref{sec:methods} outlines the study design and data collection, including the study's design details, a brief overview of LLMs, the evaluation criteria, the compilation of statistical analysis tasks, and the human evaluation rating process. Section~\ref{sec:results} presents the performance evaluation results, including the modeling and analysis of the rating scores, and summary of rater comments. Finally, Section~\ref{sec:concluding.remarks} concludes the paper by summarizing the main findings, discussing the limitations of this study, and suggesting directions for future research.

\section{Study Design and Data Collection}\label{sec:methods}
\subsection{Overview of The Study Design}

\begin{figure}
\begin{center}
\includegraphics[width=1\textwidth]{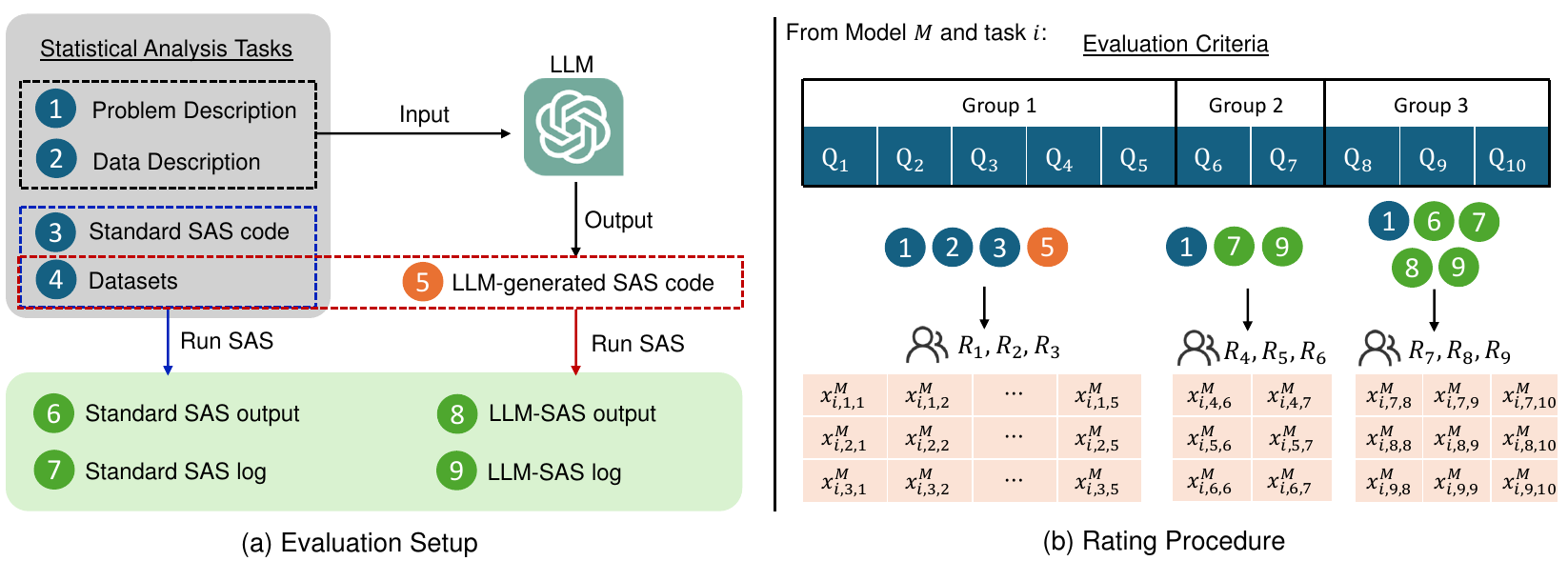}
\end{center}
\caption{Study design illustration showing the evaluation setup (a) and the rating procedure~(b). The LLMs under investigation are GPT 3.5, GPT 4.0, and Llama 3.1 70B. The standard SAS code refers to the human-verified SAS code, and those $x_{i,j,k}^{M}$'s are rating scores. }
\label{fig:flowchart}
\end{figure}

We first provide an overview of the study design under the context of  the SAS programming because of its well-structured syntax. We assess the performance of three LLMs, GPT 3.5, GPT 4.0, and Llama 3.1 70B, as introduced in Section~\ref{sec:llm.intro}. A set of 207 statistical analysis tasks was compiled, and the LLMs were tasked with generating SAS code to address these tasks. The generated code was evaluated based on three groups of criteria: code quality, executability, and output quality, detailed in Section~\ref{sec:criteria}. The study design is illustrated in Figure~\ref{fig:flowchart}, which includes the evaluation setup (a) and the rating procedure (b). Detailed explanations of the steps shown in Figure~\ref{fig:flowchart} are provided in the following sections.

\subsection{Statistical Analysis Tasks}\label{sec:stat.analysis.tasks}

To evaluate the capability of LLMs in generating code, we designed a series of statistical analysis tasks. The research team manually collected datasets and statistical analysis tasks from various online public sources. These tasks cover a wide range of basic statistical analyses across various fields, including biology, medicine, engineering, and social sciences. In total, 65 datasets (in CSV format) were gathered along with their descriptions. From these datasets, 207 statistical analysis tasks were compiled, forming a problem description and obtaining the corresponding human-verified SAS code. All collected information was systematically organized into files for efficient use and analysis.

\begin{figure}
\begin{center}
\includegraphics[width=1\textwidth]{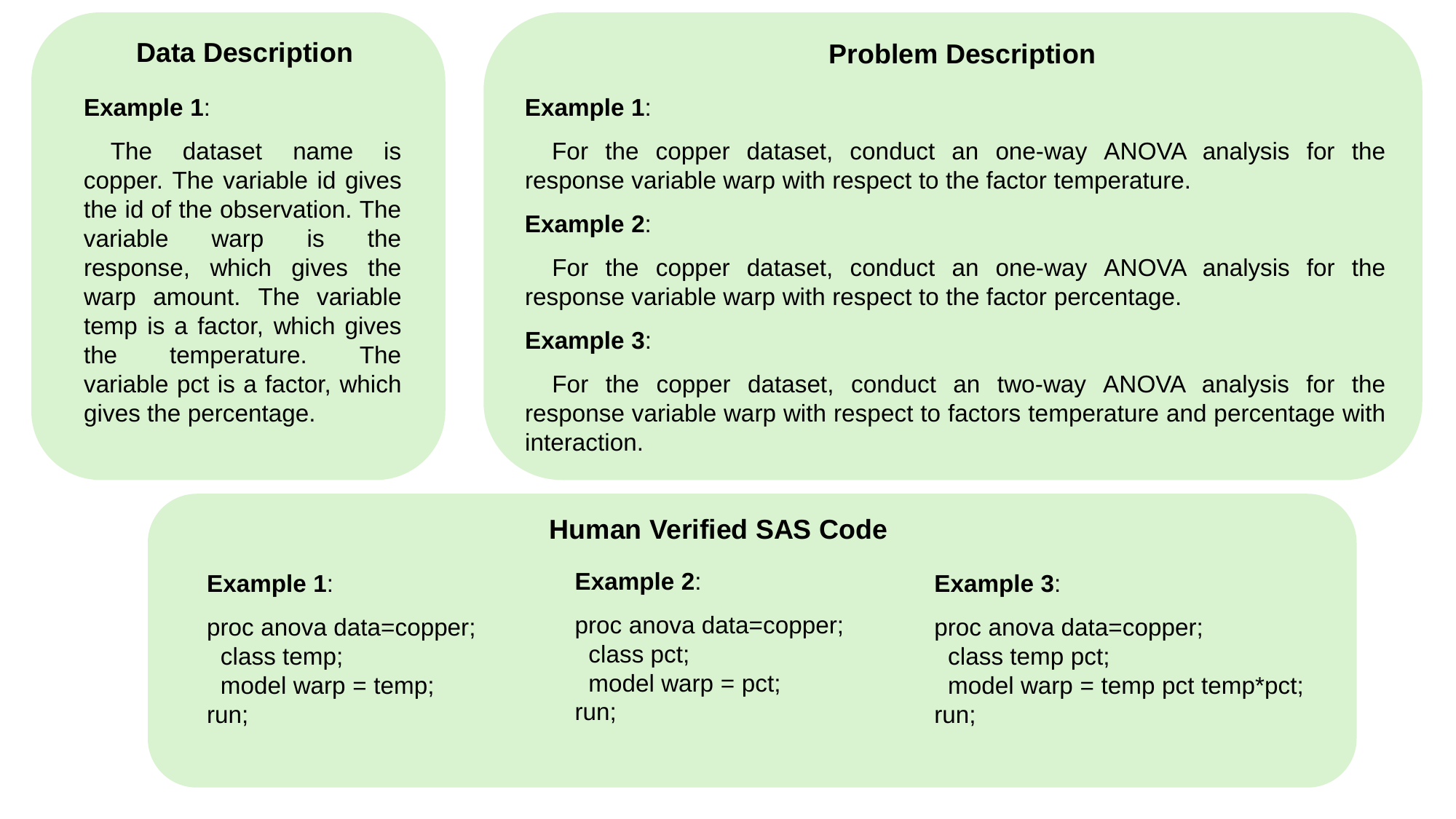}
\end{center}
\caption{Examples of data description, problem descriptions, and human-verified SAS code.}
\label{fig:DataProbelmExamples}
\end{figure}

Figure~\ref{fig:DataProbelmExamples} presents examples of data descriptions, problem statements, and human-verified SAS code. Each data description provides essential details about its corresponding dataset, including the dataset name, a general overview and background of the data, variable names, and detailed descriptions of the variables. The problem descriptions specify the statistical questions to be addressed, providing necessary information to perform the required statistical analysis tasks. The SAS code associated with each problem is verified by the researchers to ensure accuracy. Each SAS code file is tested manually to confirm it produces the correct output for the corresponding statistical problem.

Among these 207 statistical analysis tasks, their difficulty levels vary but within the undergraduate to master's level in statistics majors. These tasks cover a broad range of statistical analyses, including basic data summarization and visualization, univariate analysis, tests of means, regression and ANOVA, generalized linear models, survival analysis, and nonparametric statistics. Figure~\ref{fig:DataProbelmExamples} shows three examples of ANOVA problems that can be solved using the SAS \texttt{proc anova}. Other tasks in the set can be addressed using various SAS procedures, including \texttt{ttest},  \texttt{logistic},  \texttt{glm},  \texttt{reg},  \texttt{corr},  \texttt{npar1way},  \texttt{freq},  \texttt{univariate},  \texttt{sgscatter},  \texttt{print},  \texttt{plot},  \texttt{chart},  \texttt{means},  \texttt{phreg},  \texttt{lifetest},  \texttt{sort},  \texttt{glimmix},  \texttt{robustreg},  \texttt{sgplot},  \texttt{tabulate},  and \texttt{boxplot}. The corresponding human-verified SAS code for each task is typically under 10 lines.

\subsection{Large Language Models}\label{sec:llm.intro}

LLMs are complex deep learning networks pre-trained on vast amounts of data, enabling them to understand and generate human-like text. For each individual statistical analysis task, the LLM can generate the SAS code based on the problem description, data description, and the associated data. The evaluation procedure in this study primarily involves two versions of ChatGPT (i.e., 3.5 and 4.0) developed by OpenAI (\shortciteNP{openai2024gpt4technicalreport}), and one version of Llama (i.e., 3.1 70B) developed by Meta AI (\shortciteNP{touvron2023llamaopenefficientfoundation}). We give a brief introduction to those LLMs in the sections to follow.

\subsubsection{ChatGPT}

ChatGPT, short for Chat Generative Pre-trained Transformer, is a conversational AI system developed by OpenAI. ChatGPT can assist users with various tasks by engaging in natural language interactions. Built on a generative pre-trained transformer architecture and trained on vast amounts of data, ChatGPT can engage in dynamic conversations, answer questions, and aid in problem-solving.  In coding and programming, ChatGPT has evolved through multiple versions, each bringing improved problem-solving abilities. GPT-3.5, released in March 2022, utilizes a transformer model with dense attention layers within an LLM framework. It is particularly effective for quick code generation and debugging in various programming languages (e.g., Python, JavaScript), making it especially useful for beginners (\shortciteNP{shirafuji2023exploringrobustnesslargelanguage}). However, it struggles with more complex coding scenarios and multi-step programming tasks.

GPT-4.0, the fourth generation model in OpenAI's GPT series, was released on March 14, 2023, marking a significant advancement over its predecessors with enhanced understanding, logical reasoning, and accuracy. It is designed to handle more complex tasks and provide deeper analysis (\shortciteNP{openai2024gpt4technicalreport}). In programming, GPT-4.0 generates more reliable code for complex, multi-step tasks, code debugging, and problem-solving across diverse programming languages. Its optimized pre-trained model architecture enables smoother user interactions and improved performance. Besides the widely known GPT-3.5 and GPT-4.0 models, other versions like ChatGPT o1-preview and ChatGPT 4.0o offer unique features such as advanced reasoning and greater efficiency, providing users with diverse options for solving different problems.

\subsubsection{Llama}

The Llama series (Large Language Model Meta AI), developed by Meta AI, consists of open-source LLMs designed for various NLP tasks, including code generation. Llama 1, released in February 2023, was Meta's first open-access LLM. Its successor, Llama 2, released in July 2023, is fully open-source and widely used for specialized NLP tasks. \shortciteN{bap2024codellamaopenfoundation} introduced Code Llama, an open-source language model built on Llama 2. It outperforms previous open-source models and is competitive with proprietary models like OpenAI's Codex on several code generation benchmarks. However, challenges remain, such as complex context understanding, dependency on prompt quality (\shortciteNP{chen2021evaluatinglargelanguagemodels}), and potential security vulnerabilities (\shortciteNP{e25060888}).

Llama 3.1 is designed to enhance capabilities in multilingual understanding, coding, reasoning, and tool usage (\shortciteNP{dubey2024llama3herdmodels}). It is available in multiple versions: Llama 3.1 8B, Llama 3.1 70B, and Llama 3.1 405B, containing 8 billion, 70 billion, and 405 billion parameters, respectively. Each version is trained with different parameter counts and context lengths, impacting the quality of generated code across various tasks. Llama 3.1 8B is suitable for moderate code generation tasks, Llama 3.1 70B offers higher accuracy for more complex analytical tasks, and Llama 3.1 405B is ideal for advanced coding scenarios requiring deep contextual understanding or multi-task programming. This study focuses on evaluating the performance of Llama 3.1 70B for statistical programming tasks.

\subsection{Evaluation Criteria}\label{sec:criteria}

For the LLM-generated code, our evaluation aims to provide a comprehensive assessment on programming quality and effectiveness. Specifically, the evaluation criteria  consists of three major assessment components: Group~1, Code Correctness and Readability; Group~2, Executability; and Group~3, Output Correctness and Quality. A total of 10 questions (criteria) are designed to evaluate these components, with each criterion worths 5 points, leading to a maximum score of 50 points. The detailed questionnaire for raters is provided in Appendix~\ref{sec:criteria.details}.

Each criterion is assessed using a consistent five-point rating scale, ranging from ``Strongly Disagree'' (1) to ``Strongly Agree'' (5), ensuring a standardized approach to evaluating various aspects of the code. Zero points may also be awarded if no output is generated. This comprehensive evaluation method guarantees a systematic and adaptable process suitable for diverse coding scenarios.

The first group, code correctness and readability, focuses on code quality and it has five questions with a total of 25 points. In this category, raters examine the SAS code to assess the technical accuracy and correctness of the data step and model structure. Specific aspects such as dataset names, variable names, model options, and output options are verified for accuracy. Raters also evaluate whether the code is well-organized, logically structured, and easy to follow. Additionally, the assessment consider whether the code avoids unnecessary repetition in logic, variable declarations, and output generation, thus minimizing redundancy.

The second group, code executability,  consists of two questions worth a total of 10 points. This section evaluates whether the code generated by the LLM can execute successfully without producing errors or warning messages. If the code runs flawlessly, it automatically receives the full 10 points, bypassing further evaluation in this category.

The third group, output correctness and quality, evaluates output quality through three questions, totaling 15 points. Raters determine whether the output generated by the SAS code includes the necessary and accurate results for the statistical problem. The outputs must be clear, concise, and directly address the statistical analysis task. Additionally, the outputs should be free from duplicated results or excessive verbosity that does not contribute meaningful insights.

\subsection{Rating Scores}

As illustrated in Figure~\ref{fig:flowchart}(a), for each LLM, we provide a statistical analysis task consisting of a problem description and a data description, and request LLM to generate SAS code for the analysis. The rating scores are collected through a structured and systematic process as illustrated in Figure~\ref{fig:flowchart}(b). This process involves nine raters, denoted as $\{R_1,R_2,\dots,R_9\}$, who are selected based on their prior experience with SAS coding and statistical analysis. These raters are divided into three groups, consisting ten criteria $\{Q_1, Q_2, \dots, Q_{10}\}$ in total, with each group assigned three raters responsible for evaluating a specific aspect of the submissions. For example, raters $\{R_1,R_2,R_3\}$ are assigned to rate based on criteria $\{Q_1, Q_2, \dots, Q_{5}\}$, as illustrated in Figure~\ref{fig:flowchart}(b).

Prior to the assessment, all raters participate a comprehensive training session to have a common understanding of the evaluation criteria and consistency of the rating scale.  Since the focus of the evaluation criteria differ across groups, raters are provided with distinct sets of materials to ensure their assessments are both precise and relevant to the assigned aspects of the submissions.

For Group~1, which evaluates code correctness and readability, raters are provided with a data description, a problem description, standard SAS code (i.e., human-verified SAS code), and the SAS-generated answers from three LLMs. These materials enable raters to assess the logical structure, readability, and correctness of the code while ensuring their evaluations are grounded in the provided context and problem requirements. For Group~2, tasked with assessing executability, raters are provided with the SAS log file generated during the execution of the SAS code. The log file allows them to evaluate whether the code runs successfully without errors or warnings. By focusing solely on the execution log, Group~2 raters are able to objectively evaluate the code's ability to perform as intended. For Group~3, which evaluates output correctness and quality, raters are given the SAS log file, and the printed output file generated by the SAS code. These outputs enable them to determine whether the results align with the problem requirements, assess the clarity and relevance of the outputs, and ensure that the results are concise and free of redundancy.

Let $M_1, M_2$, and $M_3$ represent the three LLM models being evaluated. Denote the statistical analysis tasks as $T_{i},  i= 1,\ldots, 207$. For a given task $T_{i}$, the ratings assigned by rater $R_j$, on criterion $Q_k$, for the three models are denoted as:
\begin{align}\label{eqn:rating.score.notation}
\xvec_{i,j,k}=(x_{i,j,k}^{M_1},x_{i,j,k}^{M_2},x_{i,j,k}^{M_3}).
\end{align}
To minimize potential bias during the assessment, the order of the responses of each statistical analysis task from three models are randomly permuted, so raters will not know which model generates the answer. Specifically, for each task $T_i$, a permutation function $\pi_i:\{1,2,3\}\rightarrow\{1,2,3\}$ is applied, which specifies the order of each model's response after random shuffling. The anonymized ratings vector of question $T_i$ presented to rater $R_j$ on criterion $Q_k$ are then expressed as:
$$\widetilde{\xvec}_{i,j,k}=(x_{i,j,k}^{\pi_i(M_1)},x_{i,j,k}^{\pi_i(M_2)},x_{i,j,k}^{\pi_i(M_3)}).$$

Note that each criterion is rated by three raters independently. The collected ratings are submitted to a centralized scoring system, where they are aggregated, and anomalies or inconsistencies are flagged for further review. In the final stage of the evaluation process, a dedicated quality controller is assigned to each category to ensure comprehensive oversight. Following the submission of initial ratings, these quality controllers, possessing expertise in the relevant evaluation criteria, perform a detailed review of the scores to ensure accuracy and consistency. This review ensures adherence to predefined standards and resolves any inconsistencies. In the last step, the scores are unblinded and complied into data tables for subsequent analysis. In total, there are 18{,}630 scores, representing a tremendous effort in the rating process.

\section{Analysis of Rating Scores and Results}\label{sec:results}

For each answer (i.e., SAS code) generated by an LLM and for each criterion, three raters are assigned to evaluate it with a score. The average of the three scores is then calculated and used for analysis. Using the notation in \eqref{eqn:rating.score.notation}, the criterion level score is computed as,
\begin{align}\label{eqn:criterion.level.score}
\xbar_{i,k}^{M}=&\frac{1}{3}\sum_{j=1}^{3}x_{i,j,k}^{M}, \quad k=1, 2, 3, 4, 5,\\ \nonumber \xbar_{i,k}^{M}=&\frac{1}{3}\sum_{j=4}^{6}x_{i,j,k}^{M}, \quad k=6, 7,\quad \text{and}\quad
\xbar_{i,k}^{M}=\frac{1}{3}\sum_{j=7}^{9}x_{i,j,k}^{M}, \quad k=8, 9, 10,
\end{align}
for a given model $M$ and criterion $Q_k$. The group level score is computed as,
\begin{align}\label{eqn:group.score.cmpt}
x_{i,G_1}^{M}=\sum_{k=1}^{5}\xbar_{i,k}^{M}, \quad x_{i,G_2}^{M}=\sum_{k=6}^{7}\xbar_{i,k}^{M}, \quad \text{and}\quad
x_{i,G_3}^{M}=\sum_{k=8}^{10}\xbar_{i,k}^{M},
\end{align}
where $i=1, \dots, 207, M\in\{M_1, M_2, M_3\}$. The total score is computed as,
\begin{align}\label{eqn:total.score.cmpt}
x_{i}^{M}=x_{i,G_1}^{M}+ x_{i,G_2}^{M}+x_{i,G_3}^{M}, \quad i=1, \dots, 207, M\in\{M_1, M_2, M_3\}.
\end{align}
In the following sections, we will perform data analysis for the total score, group score, and criterion score, respectively. Additionally, we will assess rater variability.

\subsection{Total Score Analysis}
\begin{figure}
\begin{center}
\begin{tabular}{cc}
\includegraphics[width=0.48\textwidth]{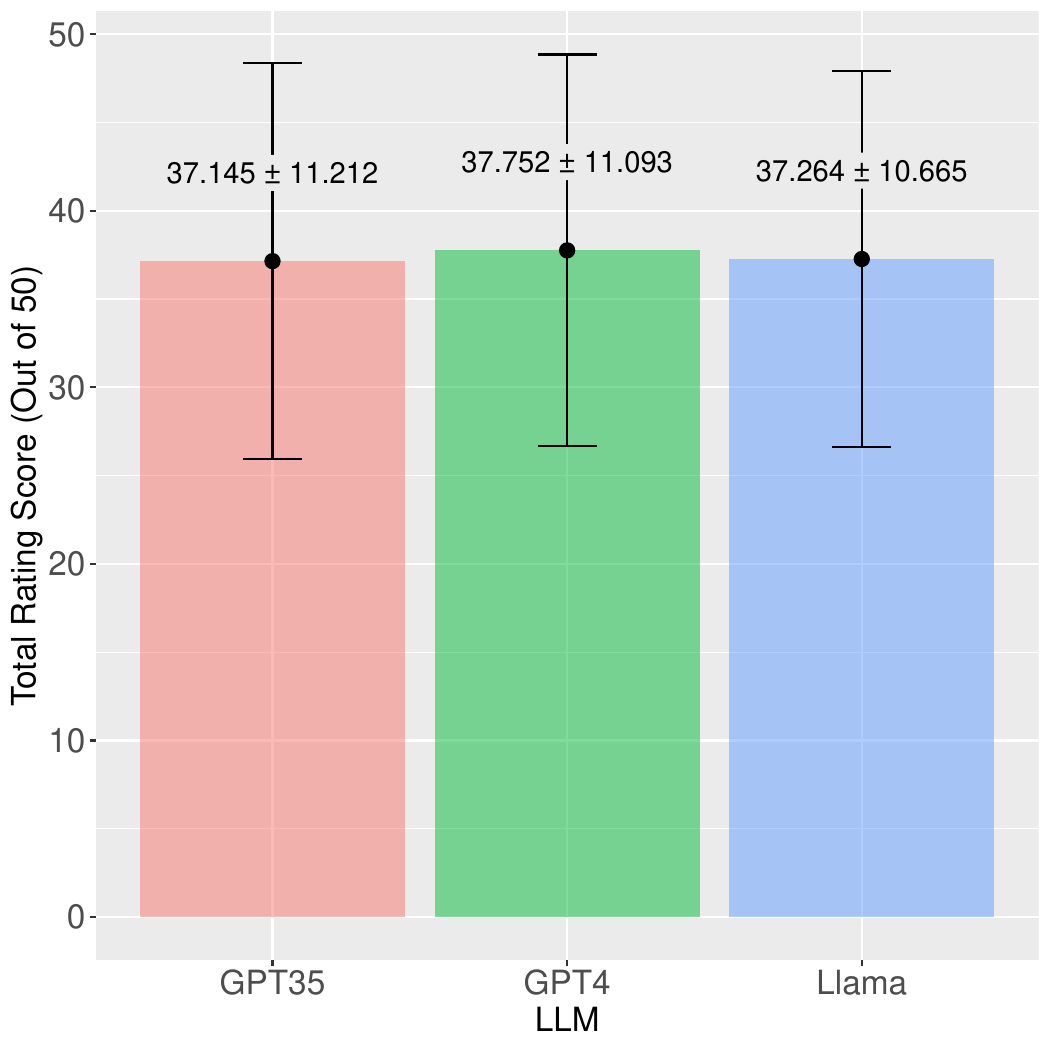}&
\includegraphics[width=0.48\textwidth]{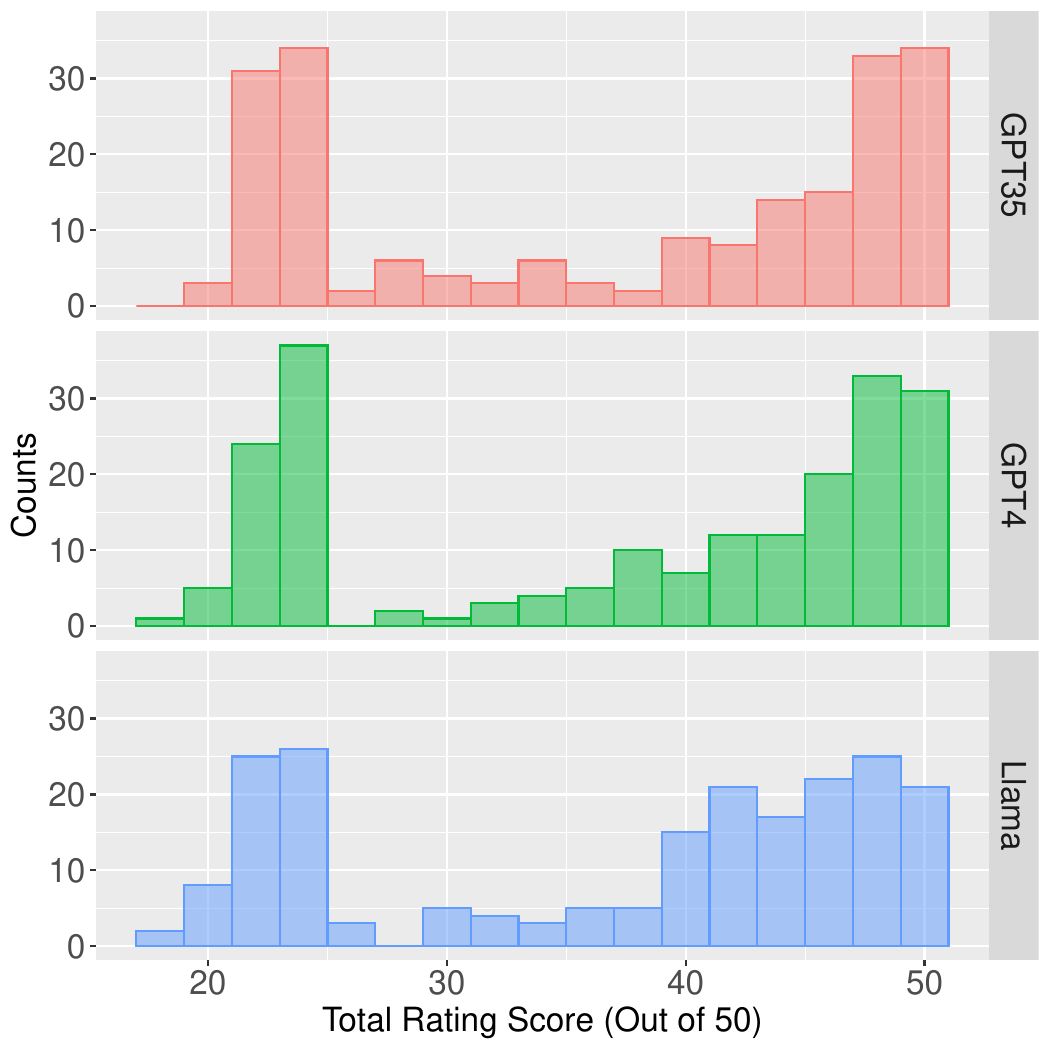}\\
(a) Mean with Error Bar & (b) Histogram
\end{tabular}
\caption{Plot of the mean total score with error bars representing the SD for the three LLMs (a), and histograms displaying the distributions of total scores (b).}\label{fig:total_score_mean_std}
\end{center}
\end{figure}

In this section, we conduct a statistical analysis of the total score to evaluate the overall performance of the LLMs in statistical programming. Using the total score defined in \eqref{eqn:total.score.cmpt}, the average total score across the three LLMs and all 207 tasks is 37.387 out of 50 (74.774\%) with a standard deviation (SD) of 10.977. Figure~\ref{fig:total_score_mean_std}(a) visualizes the mean total scores with error bars representing the SD for each of the three LLMs. The results indicate that the three LLMs perform similarly overall. To provide an intuitive perspective, if these three LLMs were students taking a statistical programming course, their total scores would correspond to a final grade of ``B'' from the instructor.

We also examine the distribution of the scores, as shown in the histograms in Figure~\ref{fig:total_score_mean_std}(b). Interestingly, a bimodal pattern emerges, which is often observed in undergraduate level statistics courses. For some tasks, the LLMs perform exceptionally well, leading to high scores, while for others, their performance is poor, resulting in low scores. This mixture of high and low performance across tasks contributes to the bimodal distribution seen in the histograms.

In the following, we consider a regression framework to formally test the difference across different LLMs. We use two dummy variables to code the LLM variable. The regression model is represented as follows:
\begin{align}\label{eqn:total.score.regression}
x_{i}^{M_l}=\beta_0+\beta_1z_{1il}+\beta_2z_{2il}+\veps_{il}, \quad i=1, \dots, 207, l=1, 2, 3,
\end{align}
where $\beta_0, \beta_1$ and $\beta_2$ are coefficients and $\veps_{il}$ is the error term. Here, $M_1, M_2, M_3$ represent the GPT 3.5, GPT 4.0, and Llama, respectively. Using model $M_1$ as the baseline, the dummy variables are defined as,
$$
z_{1il}=\left\{
                 \begin{array}{ll}
                   1, & \hbox{if $l=2$;} \\
                   0, & \hbox{otherwise.}
                 \end{array}
               \right.
\quad \text{and} \quad
z_{2il}=\left\{
                 \begin{array}{ll}
                   1, & \hbox{if $l=3$;} \\
                   0, & \hbox{otherwise.}
                 \end{array}
               \right..
$$

The regression model in \eqref{eqn:total.score.regression} can be estimated using least squares. However, since the normality assumption is not satisfied, non-parametric methods are more appropriate for statistical inference than parametric inference. These methods do not rely on distributional assumptions, providing a more reliable approach when the such assumption is violated. In this paper, we use the non-parametric bootstrap to obtain confidence intervals (CIs). By resampling the data directly, the bootstrap method effectively captures the actual variability inherent in the rating scores. We then calculate percentile-based CIs to evaluate the significance of the regression coefficients. If a CI includes zero, the effect of that LLM (e.g., GPT 4.0 or Llama 3.1 70B) compared to the baseline (GPT 3.5) is not statistically significant at the confidence level of $\alpha=0.05$. The non-parametric bootstrap algorithm is detailed in Algorithm~\ref{alg:bootstrap} in Appendix~\ref{sec:appendix.results}.

Figure~\ref{tab:total_score_boot} presents the parameter estimates of the regression coefficients and the corresponding bootstrap CIs based on the total scores, using 5{,}000 bootstrap samples. The results indicate that GPT 4.0 achieves the highest total score, followed by Llama, and then GPT 3.5. However, since the CIs for the contrasts (i.e., $M_2$ vs. $M_1$ and $M_3$ vs. $M_1$) include zero, none of these LLMs are statistically significantly better than the others.

\begin{table}
\centering
\caption{Parameter estimate of regression coefficients and the associated bootstrap CI based on the total scores (Key: $M_1$ = GPT 3.5, $M_2$ = GPT 4.0, $M_3$ = Llama). }\label{tab:total_score_boot}
\begin{tabular}{crrrc}\hline\hline
\multirow{2}{*}{Parameter}& \multirow{2}{*}{Estimate} & \multicolumn{2}{c}{95\% Bootstrap CI} &\\\cline{3-4}
&& Lower & Upper &\\\hline
Baseline ($M_1$) & 37.1449  & 35.6807 & 38.6083 &\\
$M_2$ vs. $M_1$      & 0.6071 &  $-$1.5864 & 2.7322 &\\
$M_3$ vs. $M_1$      & 0.1192 & $-$1.9694 & 2.2261 &\\
\hline\hline	
\end{tabular}
\end{table}

\subsection{Group Score Analysis}
\begin{figure}[h]
	\begin{center}
		\includegraphics[width=0.95\textwidth]{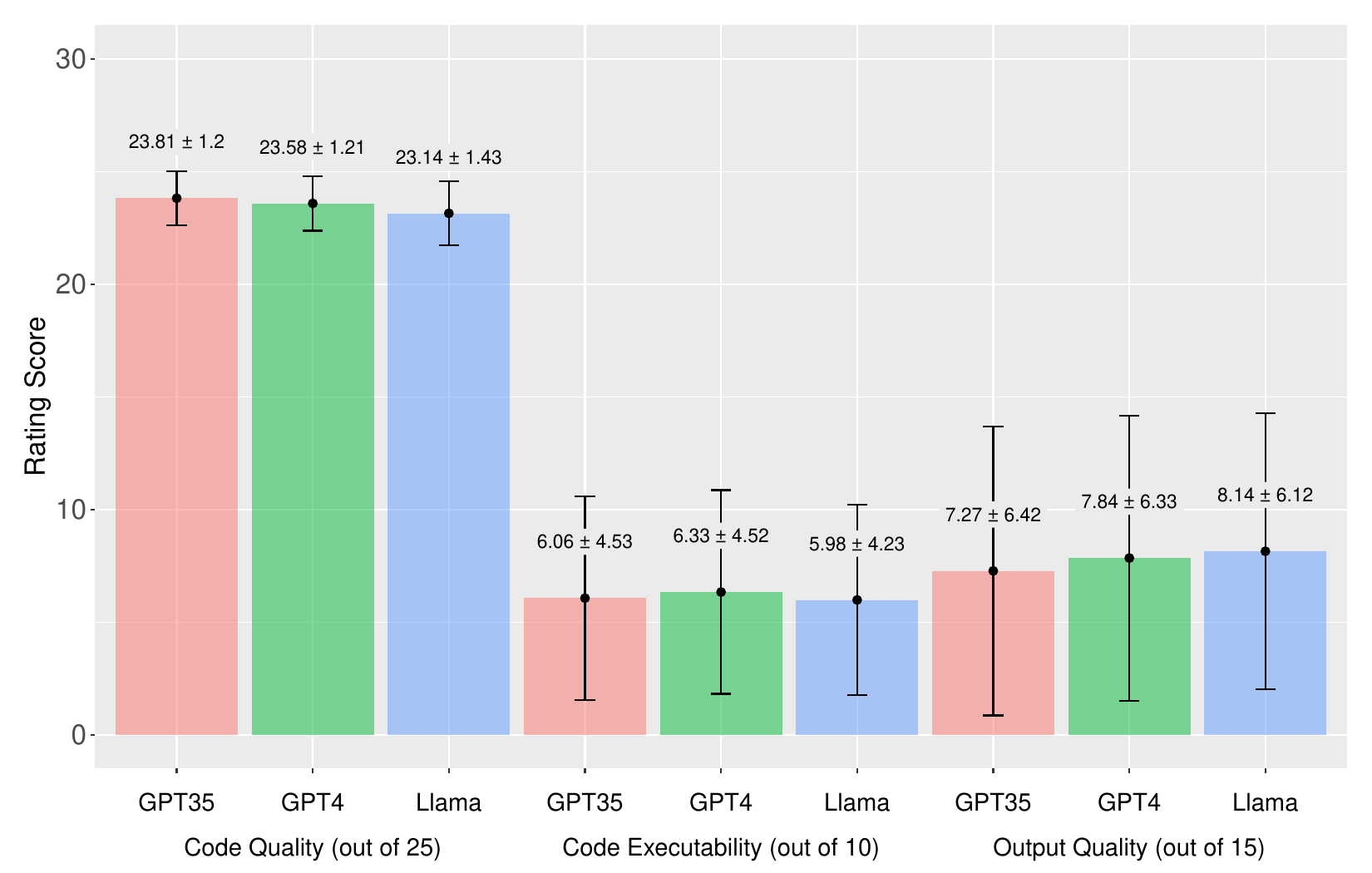}
	\caption{Plot of the mean group scores with error bars representing the SD for the three LLMs. Note that the score ranges for each group are different.}
	\label{fig:group_score_bar_LLM}
	\end{center}
\end{figure}

As mentioned earlier, the 10 rating criteria are categorized into three groups: code quality, code executability, and output quality. This section analyzes the rating scores for each group separately. In summary, for Group 1, the mean score is 23.513 out of 25 (94.052\%) with an SD of 1.309; for Group 2, the mean is 6.120 out of 10 (61.202\%) with an SD of 4.423; and for Group 3, the mean is 7.753 out of 15 (51.690\%) with an SD of 6.290. Figure~\ref{fig:group_score_bar_LLM} displays the mean group scores with error bars representing the SD. Note that the score ranges differ across the groups.

From these summary statistics, we observe that LLMs perform well in Group~1, indicating that they can generate SAS code that appears to be correct to raters most of the time, with an average score of 94.052\%. However, the average score drops significantly when the code is executed without errors (average score of 61.202\%) and is even lower when producing correct results (average score of 51.690\%).

Additionally, there is a notable increase in variability across the groups, as reflected by the SDs. The small SD in Group~1 indicates relatively consistent performance among the LLMs. In contrast, the much higher SDs in Groups~2 and~3 suggest considerable variability in performance. This indicates that for some tasks, an LLM may perform perfectly, while for others, it may fail entirely. This variability is also evident in the score distributions shown in the histograms in Figure~\ref{fig:group.hist} in Appendix~\ref{sec:appendix.results}.

We also perform formal regression modeling and conduct statistical tests. For a given group $G$, the following regression model is fitted to the group scores:
\begin{align}\label{eqn:group.score.regression}
x_{i,G}^{M_l} = \beta_0 + \beta_1 z_{1il} + \beta_2 z_{2il} + \veps_{il}, \quad i=1, \dots, 207, \; l=1, 2, 3,
\end{align}
where $ x_{i,G}^{M_l}$ represents the group score for task $i$ and model $M_l$. The group scores are computed as shown in \eqref{eqn:group.score.cmpt}, and three separate models are fitted accordingly. We use least-squares estimates along with bootstrap CIs for statistical inference.

Table~\ref{tab:boot_group_level} presents the parameter estimates of the regression coefficients and the corresponding bootstrap CIs for the group scores. An asterisk ($\ast$) indicates statistical significance at the 0.05 level. For the Group 1 score, GPT 3.5 has the highest score, followed by GPT 4.0 and Llama, with both comparisons being statistically significant. In Group 2, GPT 4.0 leads, followed by GPT 3.5 and Llama; however, neither comparison is statistically significant. For Group 3, Llama achieves the highest score, followed by GPT 4.0 and GPT 3.5, with neither comparison reaching statistical significance. Further insights into these differences will be explored in the next section, where we analyze the individual criterion scores.

\begin{table}
	\centering
	\caption{Parameter estimate of regression coefficients and the associated bootstrap CI based on group scores (Key: $M_1$ = GPT 3.5, $M_2$ = GPT 4.0, $M_3$ = Llama). An asterisk ($\ast$) indicates statistical significance at the 0.05 level.} \label{tab:boot_group_level}
\begin{tabular}{c|crrrc}\hline\hline
\multirow{2}{*}{Criterion Group} &\multirow{2}{*}{Parameter}& \multirow{2}{*}{Estimate} & \multicolumn{2}{c}{95\% Bootstrap CI} & \\\cline{4-5}
 &&& Lower & Upper & \\\hline
\multirow{3}{*}{\minitab[c]{Group 1: \\Code Quality}}&
Baseline ($M_1$) & 4.763   & 4.730& 4.795 & \\
&$M_2$ vs. $M_1$      &$-$0.047   & $-$0.092 & $-$0.000 & $\ast$ \\
&$M_3$ vs. $M_1$      & $-$0.134  & $-$0.185 & $-$0.084 & $\ast$\\\hline
\multirow{3}{*}{\minitab[c]{Group 2:\\ Code Executability}}&
Baseline ($M_1$) &  3.028  & 2.712 & 3.329 & \\
&$M_2$ vs. $M_1$      & 0.136   & $-$0.305 & 0.564  & \\
&$M_3$ vs. $M_1$      & $-$0.040  & $-$0.464 & 0.400 & \\\hline
\multirow{3}{*}{\minitab[c]{Group 3:\\ Output Quality}}&
Baseline ($M_1$) & 2.425  & 2.140 & 2.712 &\\
&$M_2$ vs. $M_1$      & 0.190  & $-$0.225 & 0.607 &\\
&$M_3$ vs. $M_1$      & 0.290  & $-$0.103 & 0.701 &\\
\hline\hline	
\end{tabular}
\end{table}

\subsection{Individual Criterion Score Analysis}

\begin{figure}
\begin{center}
\includegraphics[width=.8\textwidth]{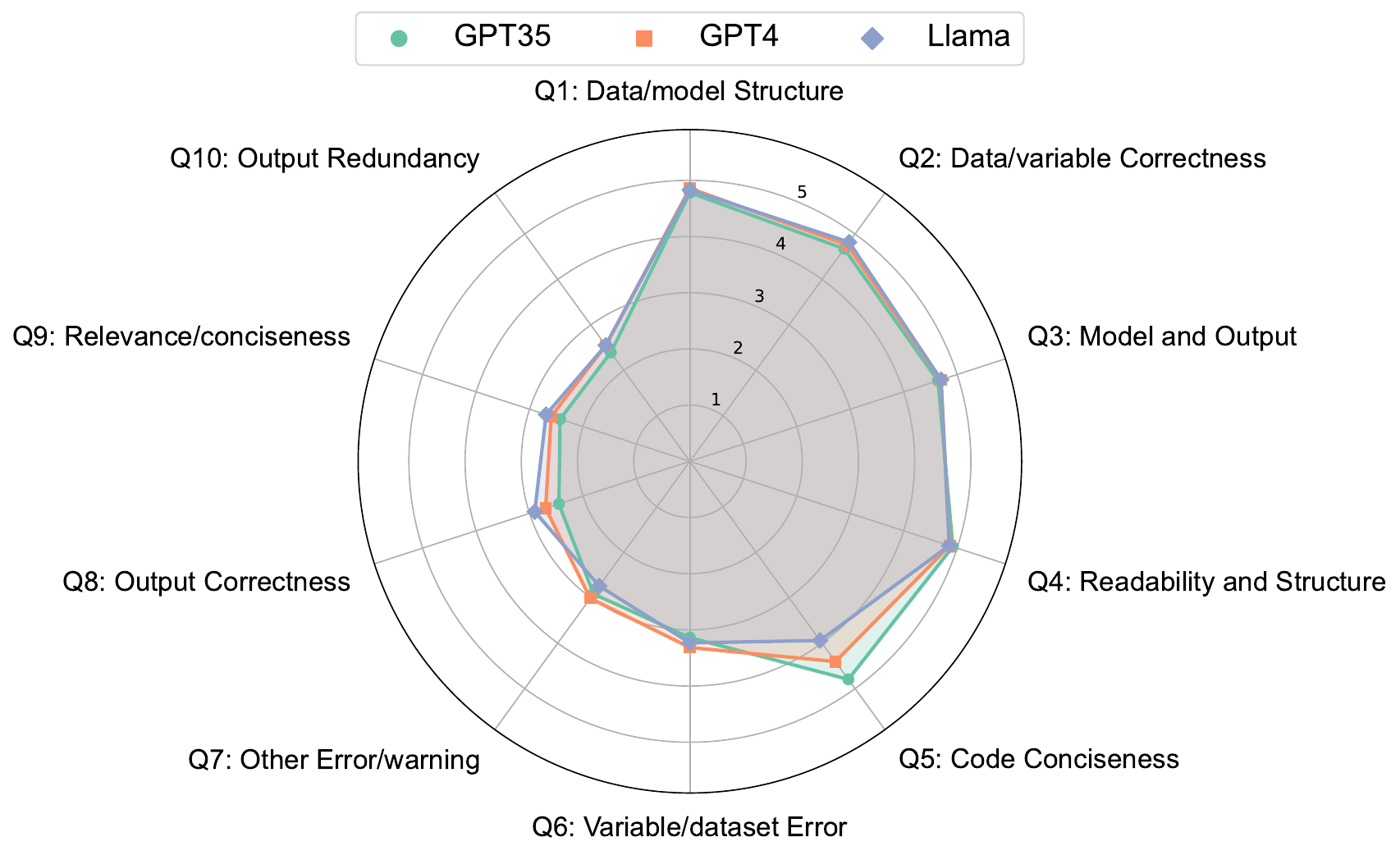}
\caption{Radar plot of the means of individual criterion scores among three LLMs.}
\label{fig:mean.criterion_score_radar_plot}
\end{center}
\end{figure}

\begin{figure}
\begin{center}
\includegraphics[width=.8\textwidth]{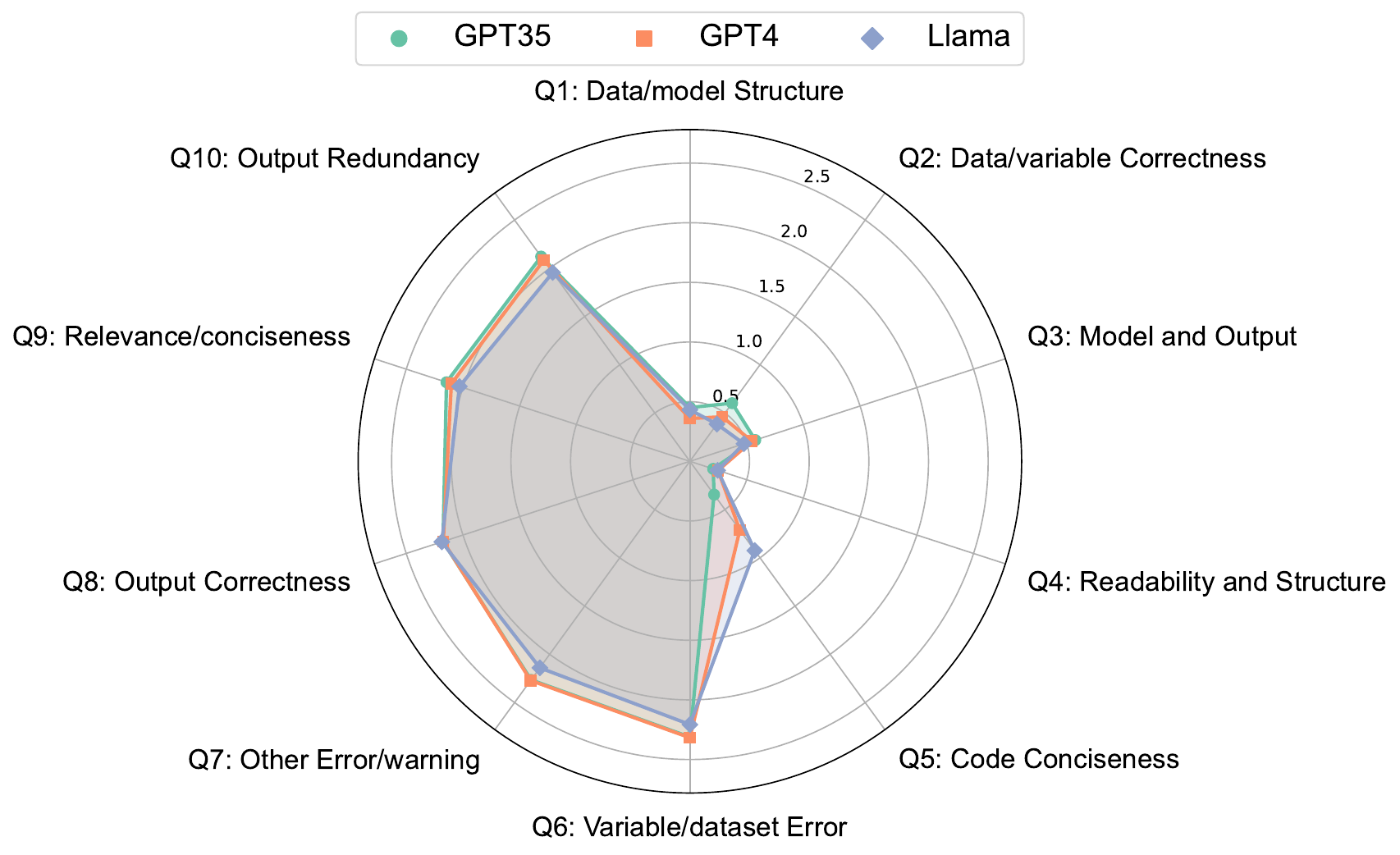}
\caption{Radar plot of the SD of individual criterion scores among three LLMs.}
\label{fig:sd.criterion_score_radar_plot}
\end{center}
\end{figure}

In this section, we examine the individual criterion scores across three LLMs. A total of 10 criteria are evaluated. Radar plots are used to visualize the mean and SD of these scores. Figure~\ref{fig:mean.criterion_score_radar_plot} displays the radar plot of the mean criterion scores for the three LLMs, while Figure~\ref{fig:sd.criterion_score_radar_plot} shows the radar plot of the SD of these scores. The distributions of the individual criterion scores are provided in Figure~\ref{fig:ind.hist} in Appendix~\ref{sec:appendix.results}.

From the mean radar plot, Q5: code conciseness and Q8: output correctness emerge as key contributors to the group and total scores. Interestingly, Llama tends to provide multiple solutions, which, while leading to redundancy, also increases its chances of delivering correct answers. For code executability, GPT 4.0 performs best in both Q6: variable/dataset errors and Q7: other errors/warnings. The SD radar plot reveals high variability in the individual criteria within Groups 2 and 3, highlighting the challenge of achieving consistent performance in executability and output quality.

Similarly, we perform formal regression modeling and conduct statistical tests for individual criterion scores. For a given criterion $k$, the following regression model is fitted to the individual criterion scores:
\begin{align}\label{eqn:individual.score.regression}
\xbar_{i,k}^{M_l} = \beta_0 + \beta_1 z_{1il} + \beta_2 z_{2il} + \veps_{il}, \quad i=1, \dots, 207, \; l=1, 2, 3,
\end{align}
where $\xbar_{i,k}^{M_l}$ represents the individual criterion score for task $i$, criterion $k$, and model $M_l$. There scores are computed as shown in \eqref{eqn:criterion.level.score}, and ten separate models are fitted accordingly. Similarly, least-squares estimates are used alongside bootstrap CIs for statistical inference. Table~\ref{tab:criterion.score.boot} shows the parameter estimates of regression coefficients and the associated bootstrap CI based on individual criterion scores. Among the significant attributes, GPT 3.5 produces the most concise and well-structured code (Q4 and Q5), while GPT 4.0 excels in data/model structure (Q1). Llama demonstrates the highest performance in data/variable correctness (Q2) and output correctness (Q8), highlighting the importance of accurate dataset and variable names for achieving correct outputs. Although not statistically significant, GPT 4.0 performs best in Q6 and Q7.

\begin{table}
\centering
\caption{Parameter estimates of regression coefficients and the associated bootstrap CI based on individual criterion scores (Key: $M_1$ = GPT 3.5, $M_2$ = GPT 4.0, $M_3$ = Llama). An asterisk ($\ast$) indicates statistical significance at the 0.05 level.}\label{tab:criterion.score.boot}

\begin{tabular}{c|rrrc|rrrc}\hline\hline
\multirow{2}{*}{Parameter}& \multirow{2}{*}{Estimate} & \multicolumn{2}{c}{95\% CI} & & \multirow{2}{*}{Estimate}
& \multicolumn{2}{c}{95\% CI} &\\\cline{3-4} \cline{7-8}
&& Lower & Upper && & Lower & Upper &\\\hline
Criterion&\multicolumn{3}{l}{Q1: Data/model Structure} & & \multicolumn{3}{l}{Q6: Variable/dataset Error}&\\\hline
Baseline ($M_1$) & 4.781 &  4.719 &	4.838  && 3.138 &  2.830  & 3.450 & \\
$M_2$ vs. $M_1$      & 0.078 & 0.003 &  0.158  &$\ast$& 0.176 & $-$0.284 & 0.610 & \\
$M_3$ vs. $M_1$      & 0.048 & $-$0.034 &0.133  && 0.093 & $-$0.341 & 0.527 & \\ \hline
Criterion&\multicolumn{3}{l}{Q2: Data/variable Correctness} & & \multicolumn{3}{l}{Q7: Other Error/warning} & \\\hline
Baseline ($M_1$) & 4.675 & 4.592 &  4.752  &&2.918 &	2.605 &	3.229 & \\
$M_2$ vs. $M_1$      & 0.082 & $-$0.020 &  0.187  &&0.097 &	$-$0.332 & 0.540 & \\
$M_3$ vs. $M_1$      & 0.147 & 0.052  &  0.244  &$\ast$&$-$0.174 &	$-$0.591 &	0.247 & \\\hline
Criterion&\multicolumn{3}{l}{Q3: Model and Output} & & \multicolumn{3}{l}{Q8: Output Correctness} & \\\hline
Baseline ($M_1$) & 4.643 &  4.562 &  4.716  && 2.451 &  2.157 & 2.750  & \\
$M_2$ vs. $M_1$      & 0.053 & $-$0.052 &  0.163 && 0.250 & $-$0.180 & 0.664  & \\
$M_3$ vs. $M_1$      & 0.058 & $-$0.042 &  0.160  && 0.452 &  0.015 & 0.870  & $\ast$\\ \hline
Criterion&\multicolumn{3}{l}{Q4: Readability and Structure} & & \multicolumn{3}{l}{Q9: Relevance/conciseness} & \\\hline
Baseline ($M_1$) & 4.919 &  4.890 &  4.946  && 2.428 &  2.1248 & 2.717  & \\
$M_2$ vs. $M_1$      &$-$0.058 & $-$0.102 & $-$0.016  &$\ast$& 0.166 & $-$0.242 & 0.570 & \\
$M_3$ vs. $M_1$      &$-$0.068 & $-$0.112 & $-$0.025  &$\ast$& 0.262 & $-$0.144 & 0.664 & \\ \hline
Criterion&\multicolumn{3}{l}{Q5: Code Conciseness} & & \multicolumn{3}{l}{Q10: Output Redundancy} & \\\hline
Baseline ($M_1$) &4.797  &  4.750 &  4.843  && 2.395 & 2.105  & 2.683 & \\
$M_2$ vs. $M_1$      &$-$0.391 & $-$0.501 & $-$0.282  &$\ast$& 0.155 & $-$0.246 & 0.565 & \\
$M_3$ vs. $M_1$      &$-$0.855 & $-$0.994 & $-$0.725  &$\ast$& 0.155 & $-$0.237 & 0.552 & \\\hline\hline
\end{tabular}
\end{table}

\subsection{Rater Variability Assessment}

As mentioned earlier, each rater blindly reviews the code and outputs from each model and assigns scores based on predefined rating rubrics. For each attribute, three raters are involved in the evaluation process independently. However, challenges arise due to potential differences in rater ``baselines'' or biases. For instance, one rater might be more lenient, consistently giving higher scores, while another could be more critical and generally assign lower scores. Therefore, it is crucial to assess rater variability to account for potential biases and ensure a fair evaluation process.

In literature, random effects models have been commonly used to assess rater variability. We use the raw score $x_{i,j,k}^{M_l}$ as shown in Figure~\ref{fig:flowchart} and consider the following random effects model,
\begin{align}\label{eqn:rater.lmm}
x_{i,j,k}^{M_l} = \beta_0 + \beta_1 z_{1ijkl} + \beta_2 z_{2ijkl} +\gamma_1 c_{1ijkl}+\cdots+\gamma_9 c_{9ijkl} +w_{j}+\veps_{ijkl}.
\end{align}
Here, we treat the LLM and individual criterion as categorical variables in the fixed effects, with coefficients $\beta$'s and $\gamma$'s and the corresponding dummy variables $z_{ijkl}$'s and $c_{ijkl}$'s. The raters are modeled as random effects, denoted by $w_j$. The random effect $w_{j}\sim\NOR(0,\sigma_{w}^2)$, while the error term $\veps_{ijkl}\sim \NOR(0, \sigma^2)$. This model allows each rater $j$ to have their own intercept, effectively capturing potential rater-to-rater variability.

The model is fitted using the R package \texttt{lme4} \shortcite{lme4R} and statistical tests were done with the R package~\texttt{lmerTest} \shortcite{lmertestR}. Table~\ref{tab:random.effect} presents the parameter estimates for the random effects model, based on all individual criterion scores from all raters. The estimated variance for the random effect is 0.0014. In comparison, the estimated variance for the error term is 2.8027. The negligible variance of the random effect indicates that rater variability is minimal compared to the variability in the error term, demonstrating a high level of consistency among the raters.

\begin{table}
\begin{center}
\caption{Parameter estimates for the random effects model based on all individual criterion scores from all raters (Key: $M_1$ = GPT 3.5, $M_2$ = GPT 4.0, $M_3$ = Llama). An asterisk ($\ast$) indicates statistical significance at the 0.05 level.}\label{tab:random.effect}
\begin{tabular}{lrrrc}\hline\hline
\multicolumn{5}{c}{Random Effects}\\\hline
Groups & Name & Variance & Std. Dev. &\\ \hline
Rater    & $\sigma_w^2$  & 0.0014  & 0.0375 & \\
Residual & $\sigma^2$    & 2.8027  & 1.6741 &  \\
\multicolumn{5}{c}{\texttt{Number of obs: 18630, groups:  Rater, 9}}\\\hline\hline
\multicolumn{5}{c}{Fixed effects}\\\hline
Parameter & Estimate & Std. Error & $t$ value &\\\hline
Baseline ($M_1$)  & 4.7991  & 0.0476  & 100.624 & \\
$M_2$ vs. $M_1$  & 0.0607  & 0.0300  & 2.021  &  $\ast$ \\
$M_3$ vs. $M_1$   & 0.0119  & 0.0300  & 0.397 &   \\
\multicolumn{5}{c}{\texttt{Output for individual criterion omitted.}}\\\hline\hline
\end{tabular}
\end{center}
\end{table}

\subsection{Rater Comments and Areas for Improvement}

During the rating process, each rater was asked to summarize the problems they encountered. These insights provide valuable feedback on areas where LLMs can be improved. For Group~1, LLMs sometimes produce excessive code by offering multiple solutions for the same problem or including unnecessary steps beyond the problem's scope. This issue is particularly noticeable in Llama. The generated code is occasionally overly complex and poorly organized, using unnecessarily complicated proc steps. Some SAS code includes incorrect model or output options, leading to incompatibility with proc steps or failure to produce the required results. Models sometimes use incorrect dataset and variable names. Although rare, the models occasionally misunderstand the problem entirely, generating incorrect code.

For Group~2, the following issues were observed. Running errors are often due to errors in dataset and variable names. Syntax errors include unrecognized arguments, missing operators (e.g., `-', `*', `:'), and incorrect keywords, such as using ``residuals'' instead of ``residual.'' The code sometimes references non-existent variables, such as ``some\_coll'' when no such variable exists in the dataset. Certain procedures, such as proc logistic, require specific variable types, like categorical variables, but the provided data may not meet these requirements. Errors can also arise from incomplete data imports or missing data sheets needed for the analysis. For Group~3, the main problem is a lack of relevance and conciseness. The outputs often miss important details or include unnecessary information. In some cases, the models failed to provide the necessary results.

For future improvements, LLMs should be trained to enhance accuracy in extracting the correct datasets and variables and to improve their syntax understanding of SAS. Incorporating statistical knowledge is crucial for accurate model specification. It is also important for LLMs to generate the appropriate amount of code needed to solve the problem without unnecessary complexity. From the user's perspective, providing more specific problem instructions could improve performance; however, this would require more human inputs.

\section{Conclusions}\label{sec:concluding.remarks}

In this paper, we propose a systematic framework to evaluate the performance of LLMs in statistical programming, with a focus on SAS programming. We developed a comprehensive set of evaluation criteria and invested significant human effort in the rating process. Our analysis provides an overall assessment of LLM performance in statistical programming with SAS. Although LLMs can generate code that appears correct, there is still considerable room for improvement, particularly in code executability and output accuracy. No single LLM consistently outperformed the others in overall ratings, but each model demonstrated strengths in certain criteria and weaknesses in others. We also identified potential areas for LLM improvement that could eventually enhance automatic statistical analysis. The data we collected on statistical analysis tasks and rating scores provide a valuable data infrastructure for future research aimed at evaluating and improving LLM performance in statistical programming.

Our research has several limitations. First, our data collection methodology may introduce selection bias, as datasets were manually collected from online sources. Although we collected 207 statistical problems for diversity, they do not fully represent all types of statistical analyses. Future research should adopt more structured data collection methods to ensure broader coverage. The evaluation process also has limitations due to the reliance on human judgment, which introduces subjectivity. Differences in how evaluators interpret rating questions and scales could impact consistency. Additionally, while our evaluation criteria were carefully designed, they might not capture all aspects of code correctness and effectiveness. Finally, our study focused on relatively simple SAS code problems, reflecting common real-world scenarios but not fully testing the LLMs' capabilities with complex statistical analyses. This limited our ability to assess the models' performance on more advanced programming tasks.

There are some interesting areas for future research. First, we utilize pre-set weights to integrate criteria from various perspectives. This approach allows for a straightforward combination of factors, but it also invites future exploration.  For instance, methods such as principal component analysis could be employed to derive weights that better capture the underlying structure and relationships among criteria, potentially improving overall performance and interpretability. Second, while this study focused on SAS programming, it will be also interesting and important to study the performance of LLM in R programming. The availability of data is important for statistical research in artificial intelligence (\shortciteNP{Zhengetal2025-datareview}). Data serves as the foundation for evaluating and developing models, as well as for assessing the effectiveness of different methodologies. The dataset curated in this paper provides a valuable resource for benchmarking existing metrics. For example, \shortciteN{song2024comprehensive} conducted an extensive comparison of metrics commonly employed in machine learning applications. Similarly, our dataset could be instrumental in performing analogous evaluations for coding metrics in the domain of statistical programming. Furthermore, the dataset has potential applications in the development of improved automated metrics tailored to statistical programming. This represents a critical step toward enabling automatic statistical analysis and lead to the third generation of statistical software (e.g., \shortciteNP{Minetal2024-appliedstat}), in which users interact with the software using natural languages. By leveraging our dataset, researchers can design and validate metrics that facilitate robust, automated evaluation processes, thereby advancing the field of statistical programming and its integration with AI.

%
%

\section*{Data Availability Statement}
The data used in this paper are available at GitHub repository: \url{https://github.com/yili-hong/StatLLM}.

\section*{Acknowledgments}

The authors acknowledge the Advanced Research Computing program at Virginia Tech for providing computational resources. The work by Deng and Hong was supported in part by the COS Dean's Discovery Fund at Virginia Tech (Award: 452021). The work by Hong was supported in part by the Data Science Faculty Fellowship (Award: 452118) at Virginia Tech.

\appendix
\section{Detailed Evaluation Criteria}\label{sec:criteria.details}

The following gives the detailed evaluation criteria.

\def\baselinestretch{0.8}
{\small
\subsection*{\small Group 1: Code Correctness and Readability (25 Points)}
If no codes are generated, all the following criteria receive zero points.
\begin{itemize}
\item[Q1:]	The data step/model proc and model structure are correct.\\
(Rating Scale: 1 = Strongly Disagree; 2= Disagree; 3 = Neutral; 4 = Agree; 5 = Strongly Agree.)
\item[Q2:]	The dataset names and variables are correct.\\
(Rating Scale: 1 = Strongly Disagree; 2= Disagree; 3 = Neutral; 4 = Agree; 5 = Strongly Agree.)
\item[Q3:]	Model options and output options are correct.\\
(Rating Scale: 1 = Strongly Disagree; 2= Disagree; 3 = Neutral; 4 = Agree; 5 = Strongly Agree.)
\item[Q4:]	Readability and structure: Code is well-organized, logically structured, and easy to follow.\\
(Rating Scale: 1 = Strongly Disagree; 2= Disagree; 3 = Neutral; 4 = Agree; 5 = Strongly Agree.)
\item[Q5:] 	Conciseness: Code avoids unnecessary repetition in logic, variable declarations, and output generation, minimizing redundancy.\\
(Rating Scale: 1 = Strongly Disagree; 2= Disagree; 3 = Neutral; 4 = Agree; 5 = Strongly Agree.)
\end{itemize}

\subsection*{\small Group 2: Executability (10 points)}
If the code can be executed successfully without error or warning messages, we assign 10 points and skip criteria Q6 and Q7. If there is no output and the absence of output is due to an error in the log files, we assign zero points. However, if there is no output but no error in the log files, indicating that the code executed successfully without producing a result, we evaluate it based on the error and warnings generated.

\begin{itemize}
\item[Q6:] Error due to variable and dataset names: The log file does not contain errors on variable name and dataset names.\\
(Rating Scale: 1 = Strongly Disagree; 2= Disagree; 3 = Neutral; 4 = Agree; 5 = Strongly Agree.)
\item[Q7:] Other error and warning messages: The log file is free of other error and warning messages other than those in Q6. For example, syntax errors, model estimation errors, convergence warnings.\\
(Rating Scale: 1 = Strongly Disagree; 2= Disagree; 3 = Neutral; 4 = Agree; 5 = Strongly Agree.)
\end{itemize}

\subsection*{\small Group 3: Output Correctness and Quality (15 points)}
If no outputs are generated, all the following criteria receive zero points.
\begin{itemize}
\item[Q8:] Necessary and correct output: Outputs contains necessary and correct results for the statistical problem.\\
(Rating Scale: 1 = Strongly Disagree; 2= Disagree; 3 = Neutral; 4 = Agree; 5 = Strongly Agree.)
\item[Q9:] Relevance and conciseness: Results are must clear, concise, and relevant to answer the question.\\
(Rating Scale: 1 = Strongly Disagree; 2= Disagree; 3 = Neutral; 4 = Agree; 5 = Strongly Agree.)
\item[Q10:] Redundancy: Outputs do not contain duplicate results or excessive verbosity that does not add value.\\
(Rating Scale: 1 = Strongly Disagree; 2= Disagree; 3 = Neutral; 4 = Agree; 5 = Strongly Agree.)
\end{itemize}
}
\def\baselinestretch{1.25}

\section{Additional Details and Results}\label{sec:appendix.results}

\begin{figure}
	\begin{center}
		\begin{tabular}{ccc}
			\includegraphics[width=.31\textwidth]{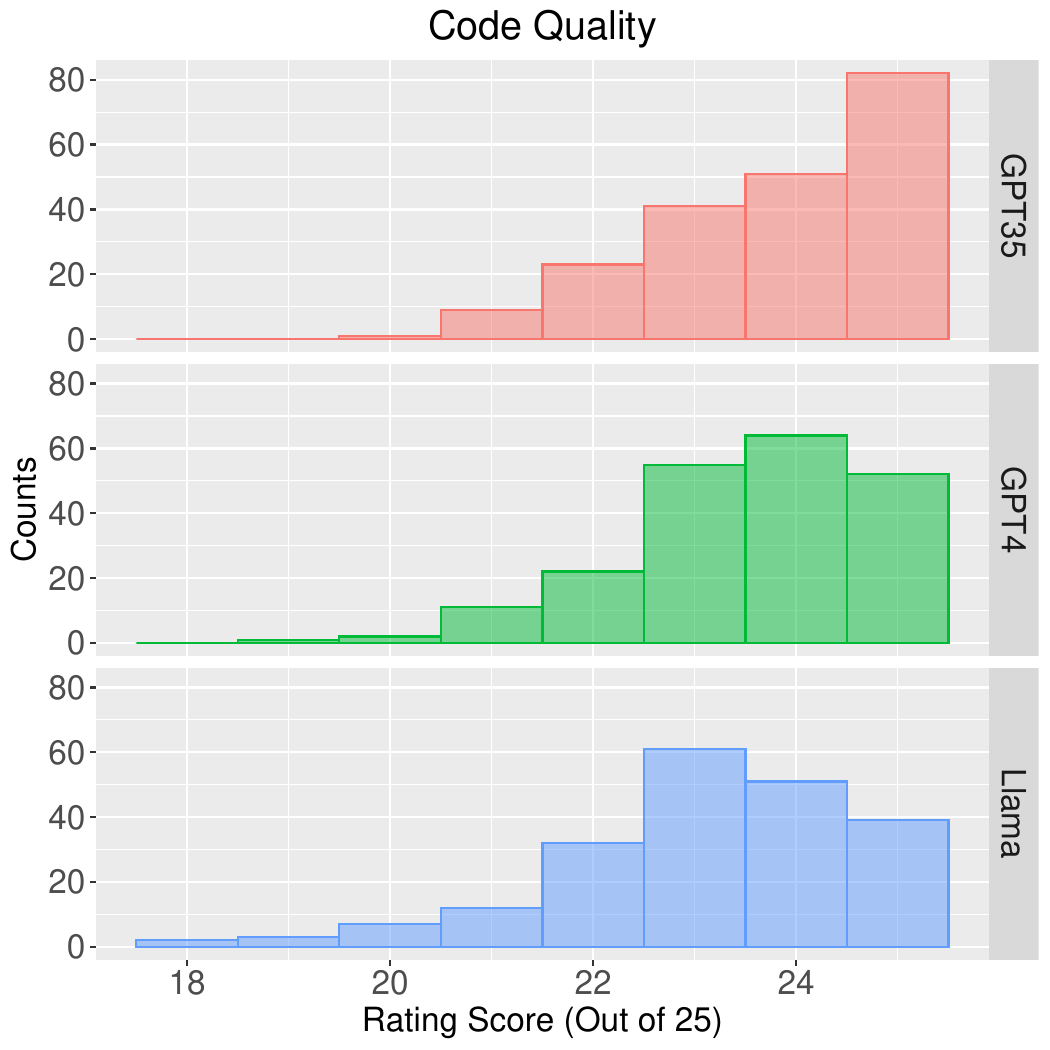} &
			\includegraphics[width=.31\textwidth]{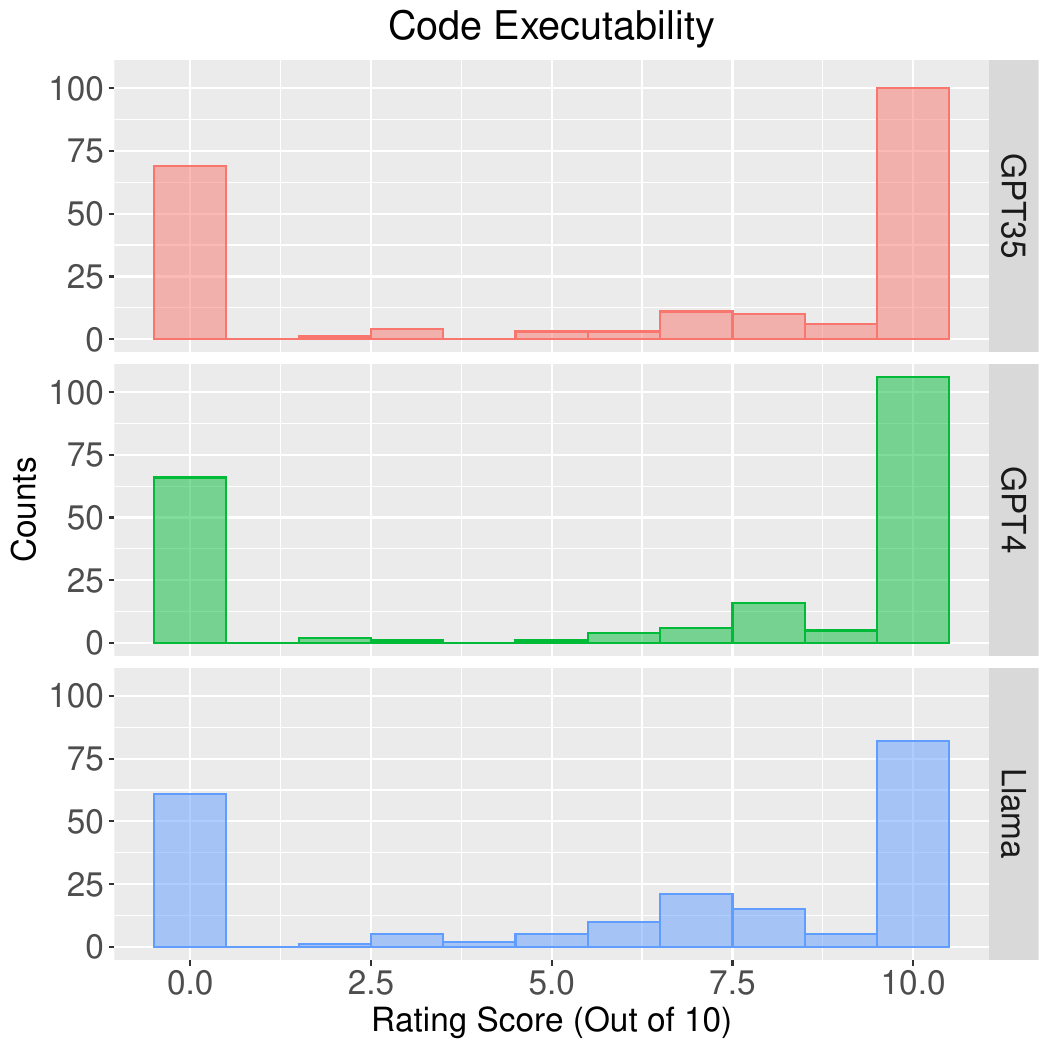} &
			\includegraphics[width=.31\textwidth]{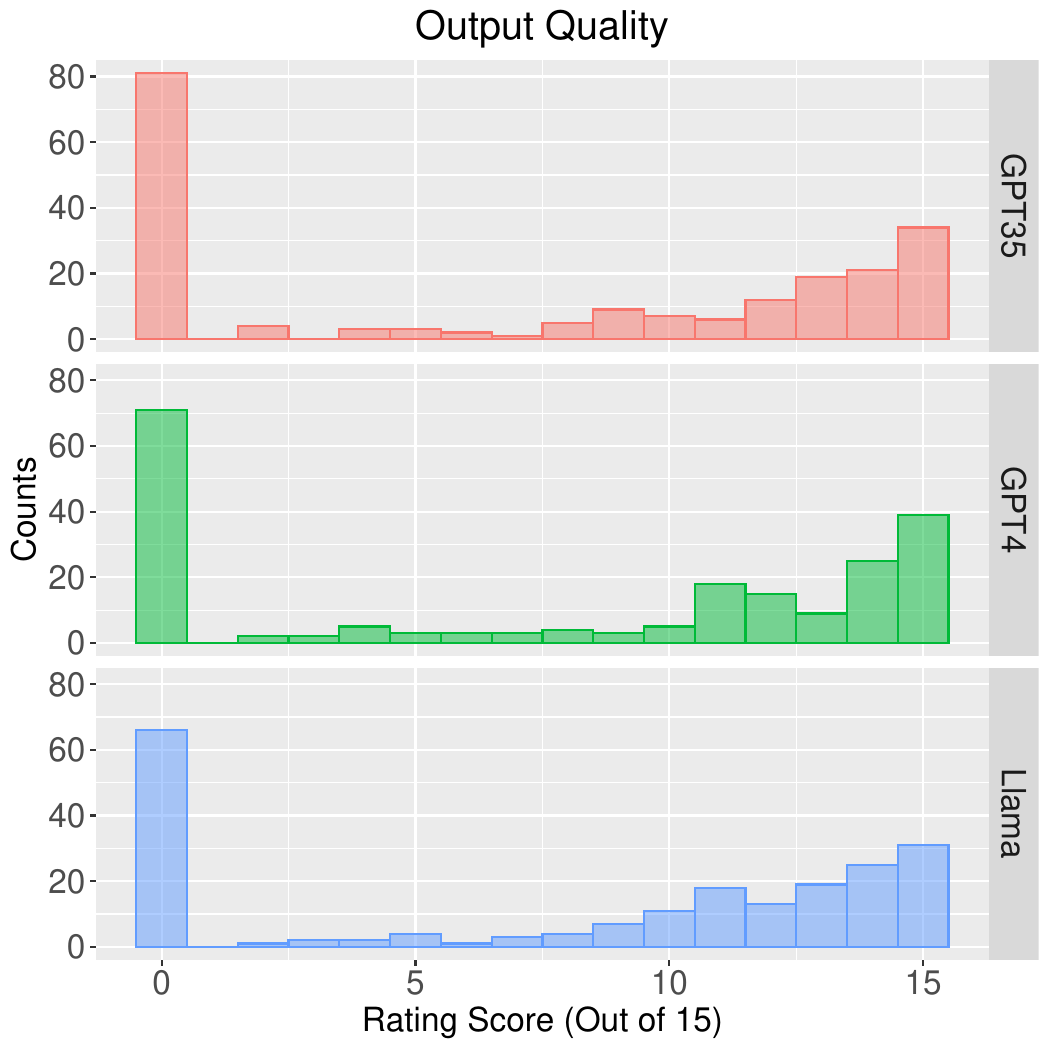}
		\end{tabular}
	\caption{Histograms of group scores (i.e., code quality, code executability, and output quality) across three LLMs.}
	\label{fig:group.hist}
	\end{center}
\end{figure}

\begin{figure}
	\begin{center}
		\begin{tabular}{ccc}
			\includegraphics[width=.31\textwidth]{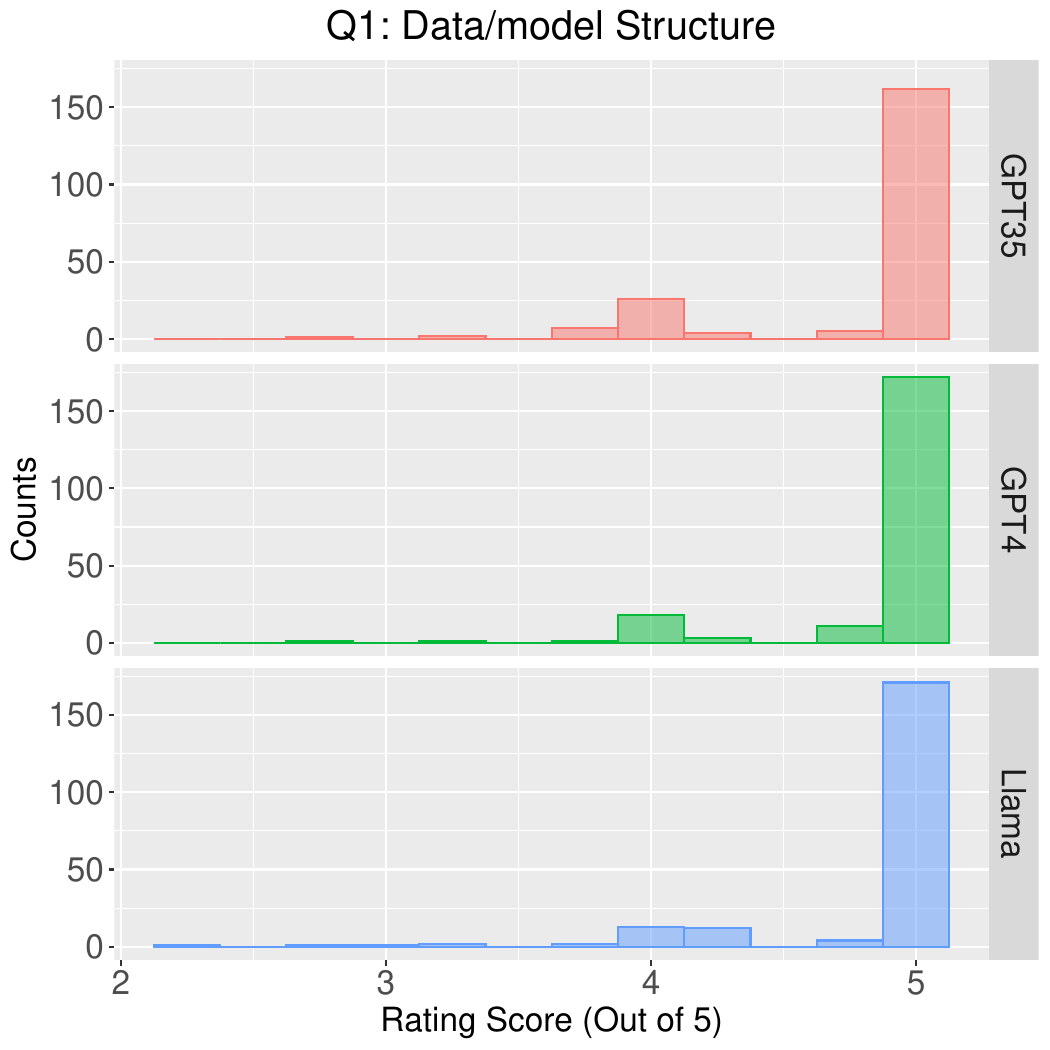} &
			\includegraphics[width=.31\textwidth]{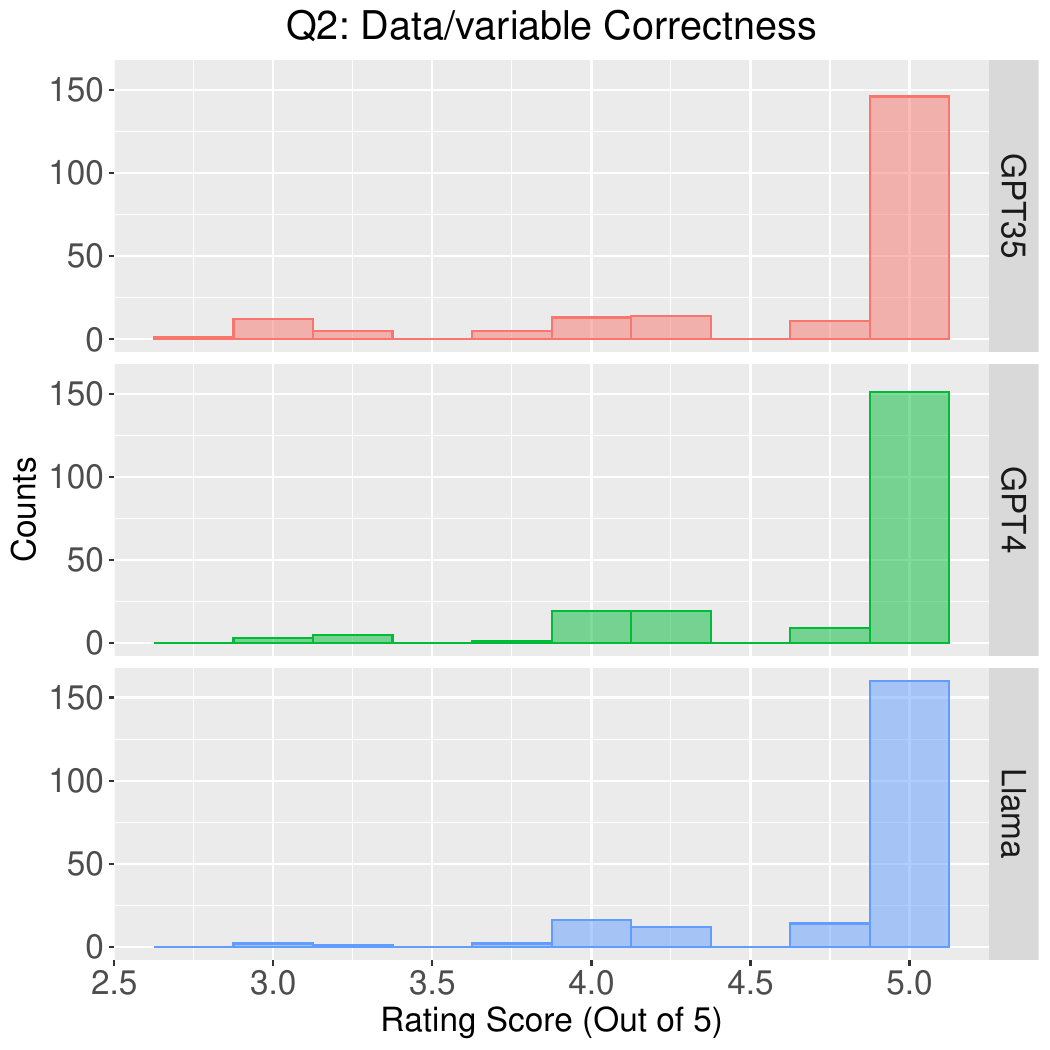} &
			\includegraphics[width=.31\textwidth]{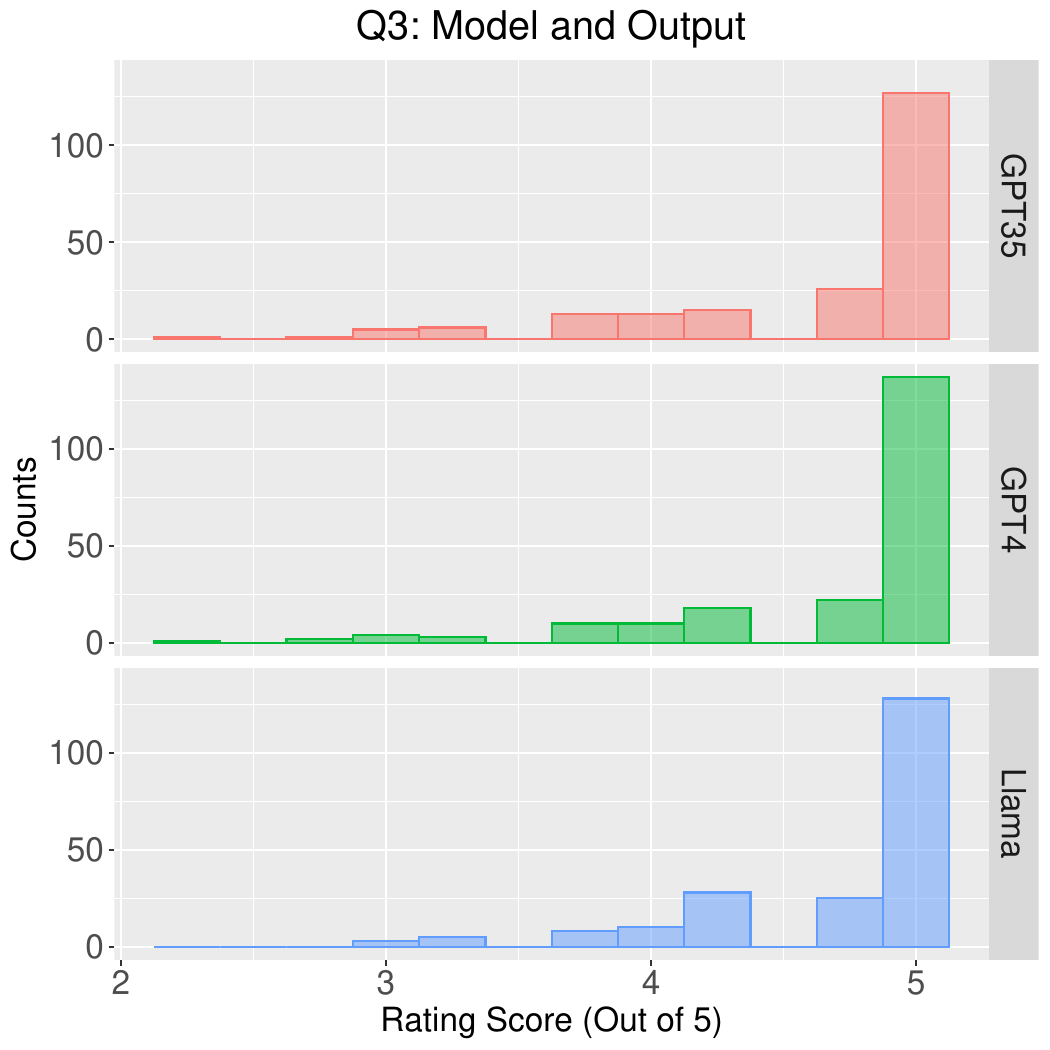} \\
			\includegraphics[width=.31\textwidth]{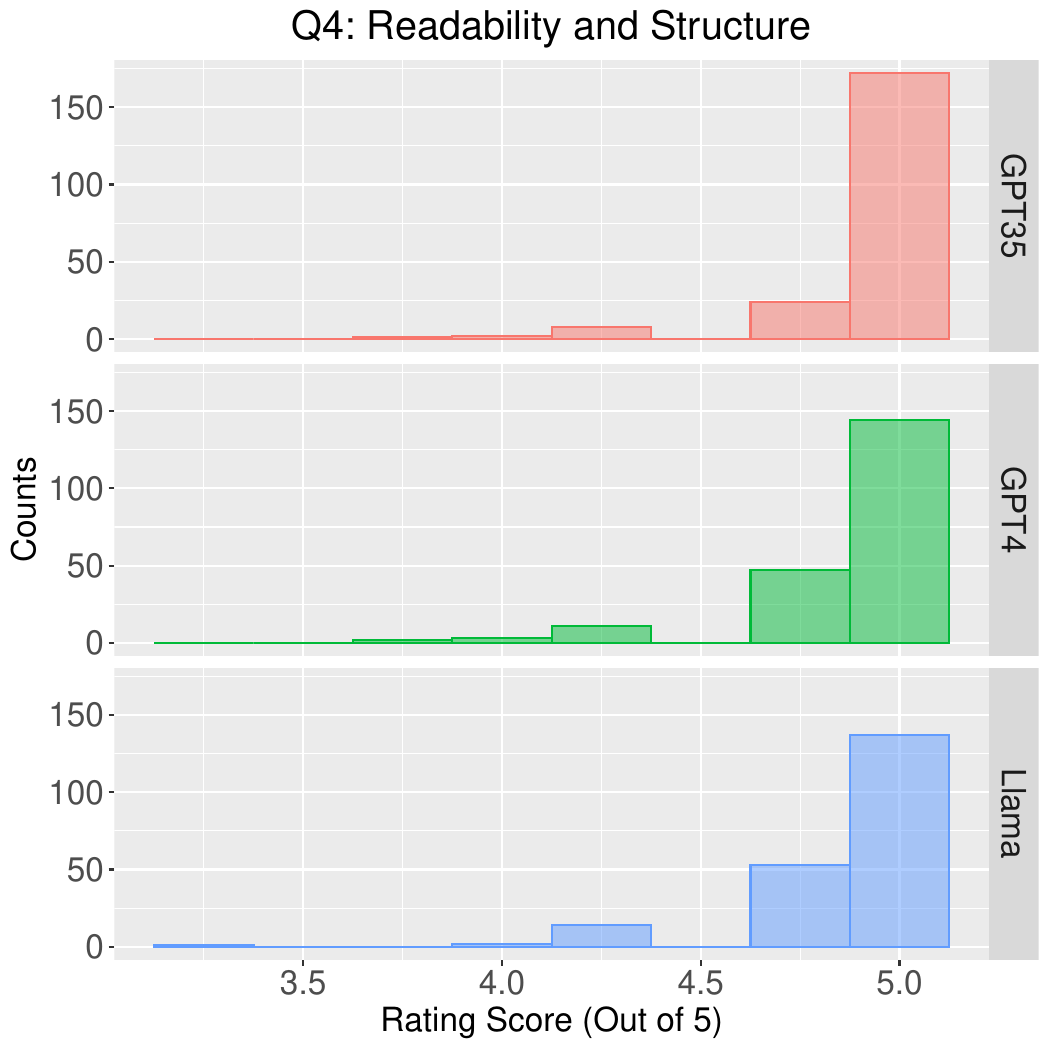} &
			\includegraphics[width=.31\textwidth]{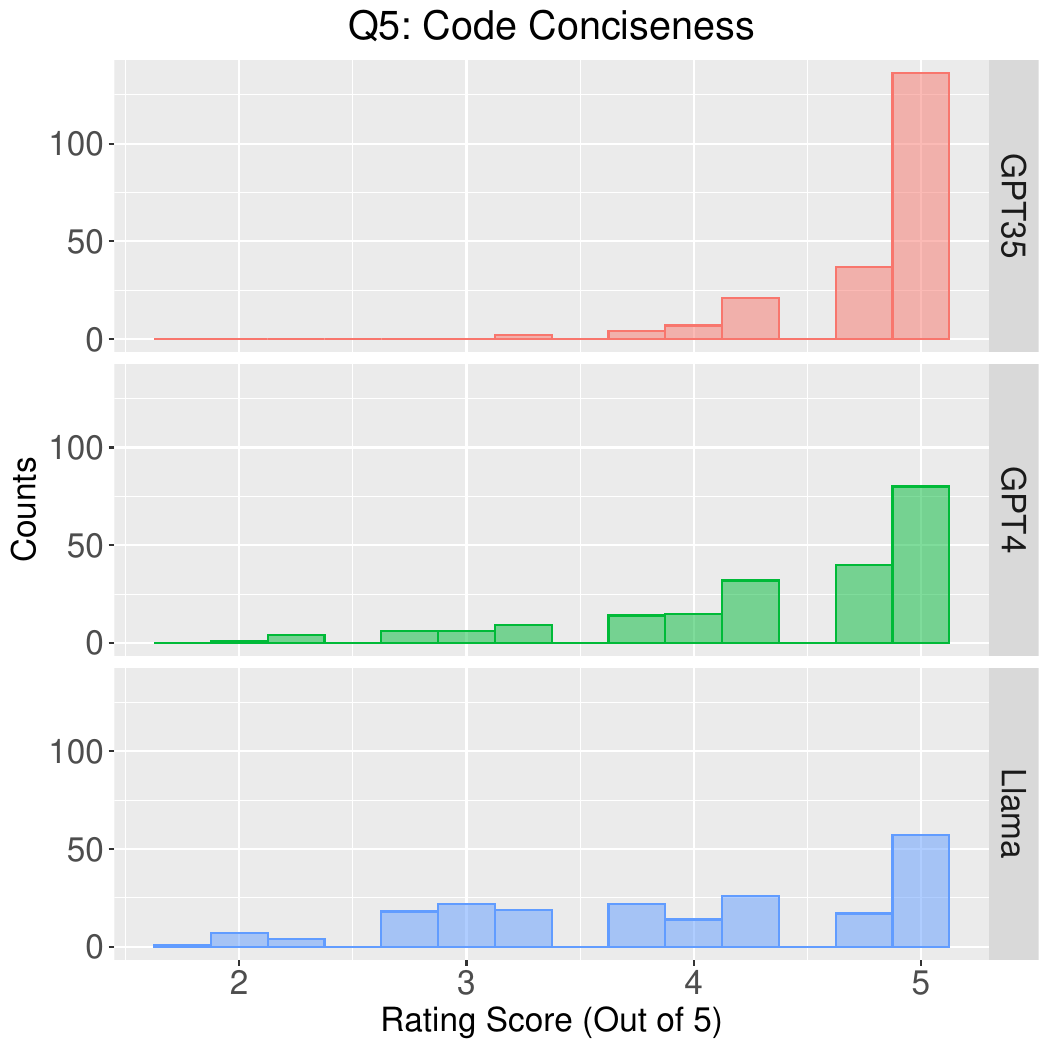} &
			\includegraphics[width=.31\textwidth]{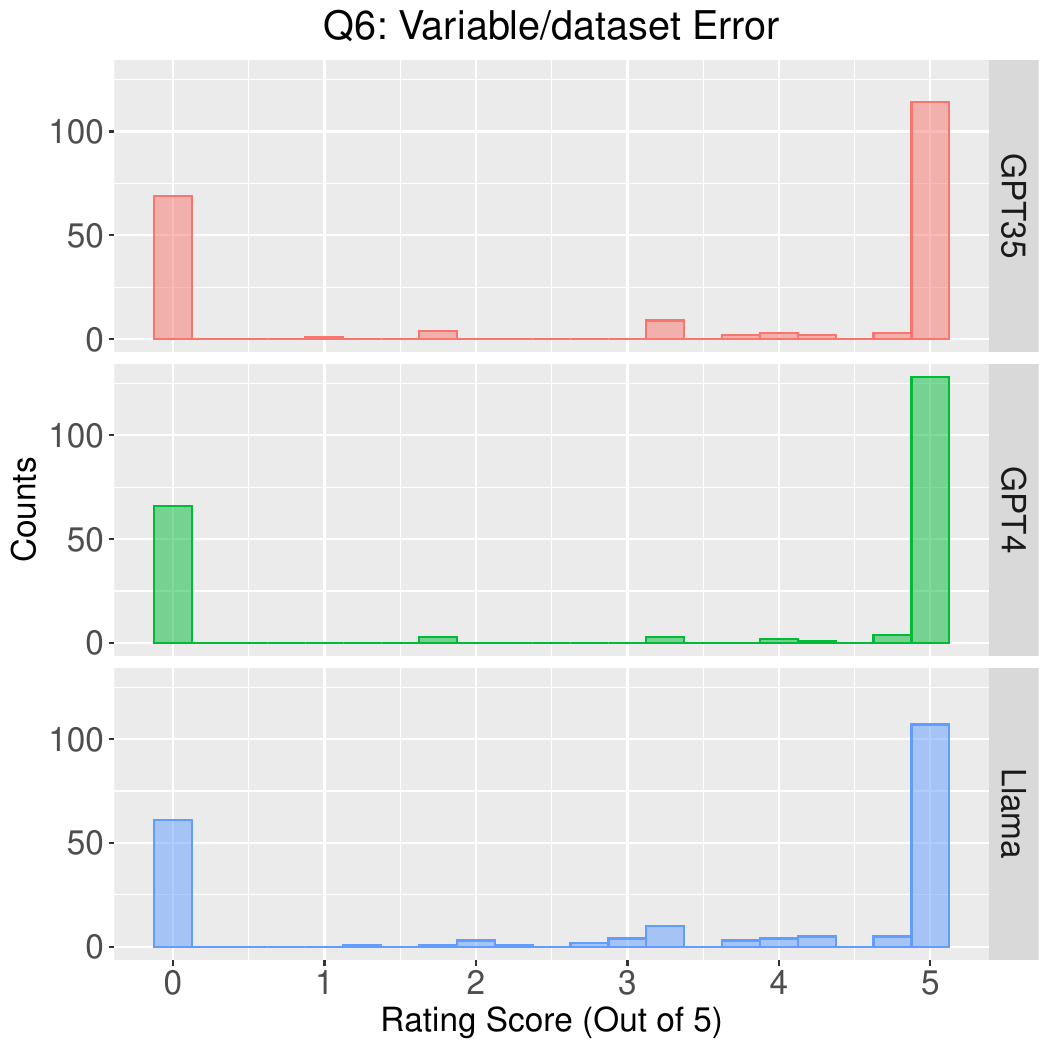} \\
			\includegraphics[width=.31\textwidth]{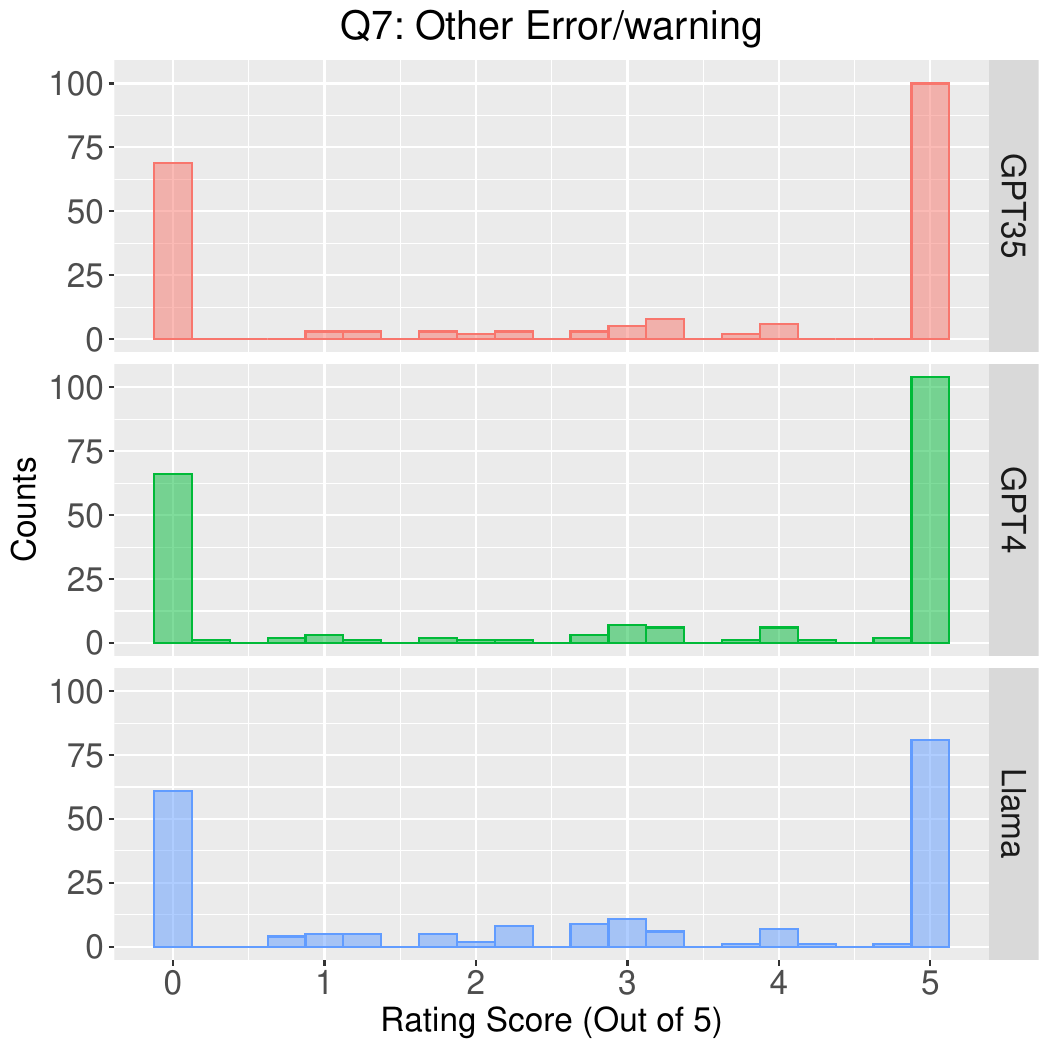} &
			\includegraphics[width=.31\textwidth]{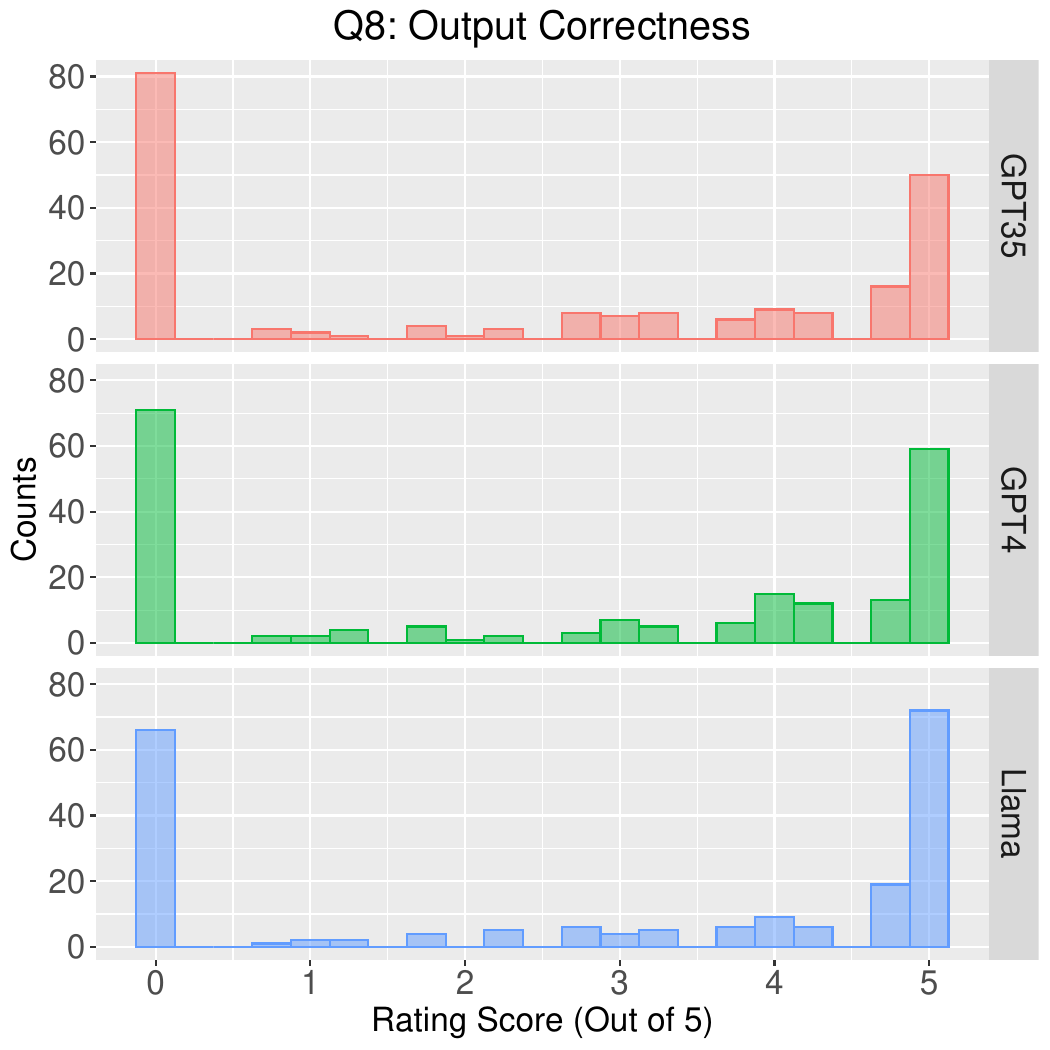} &
			\includegraphics[width=.31\textwidth]{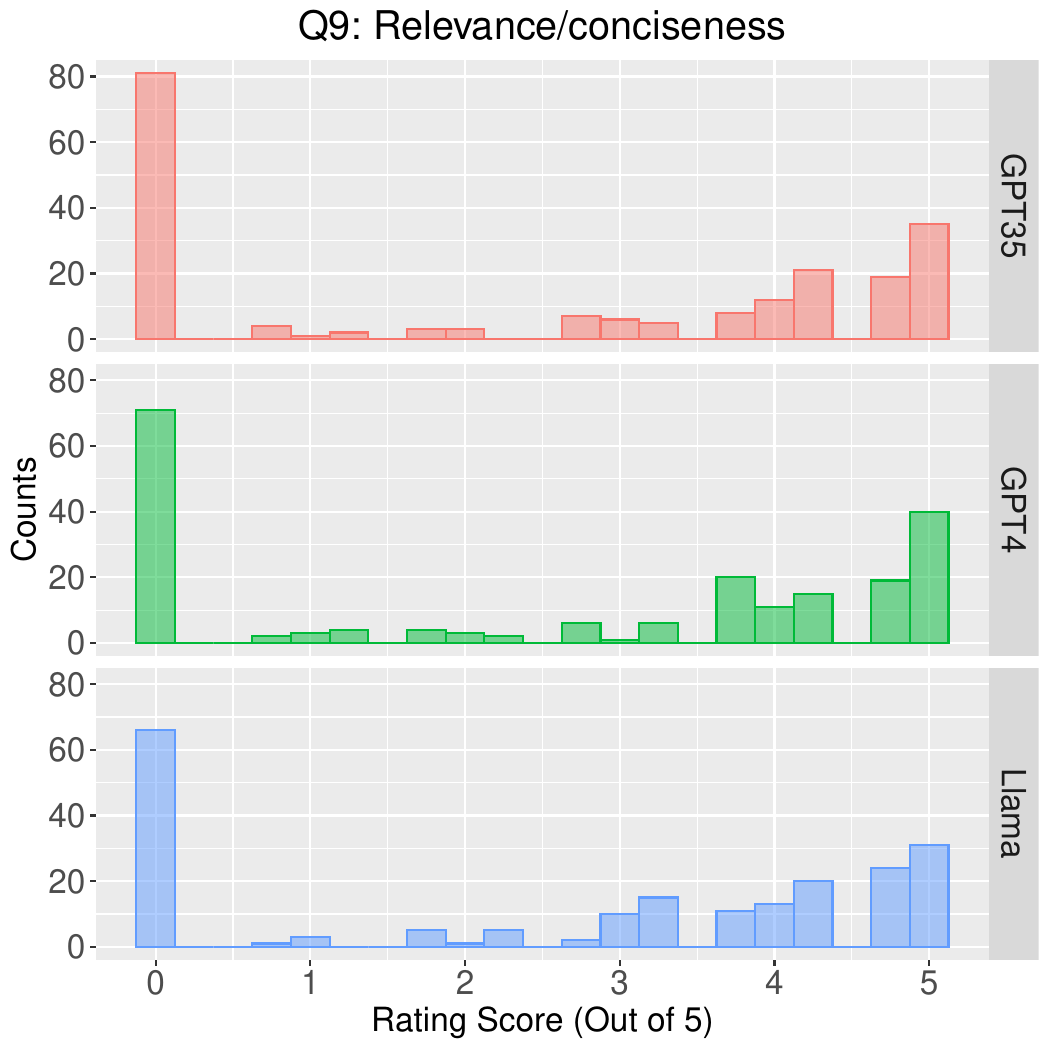} \\
			 &
			\includegraphics[width=.31\textwidth]{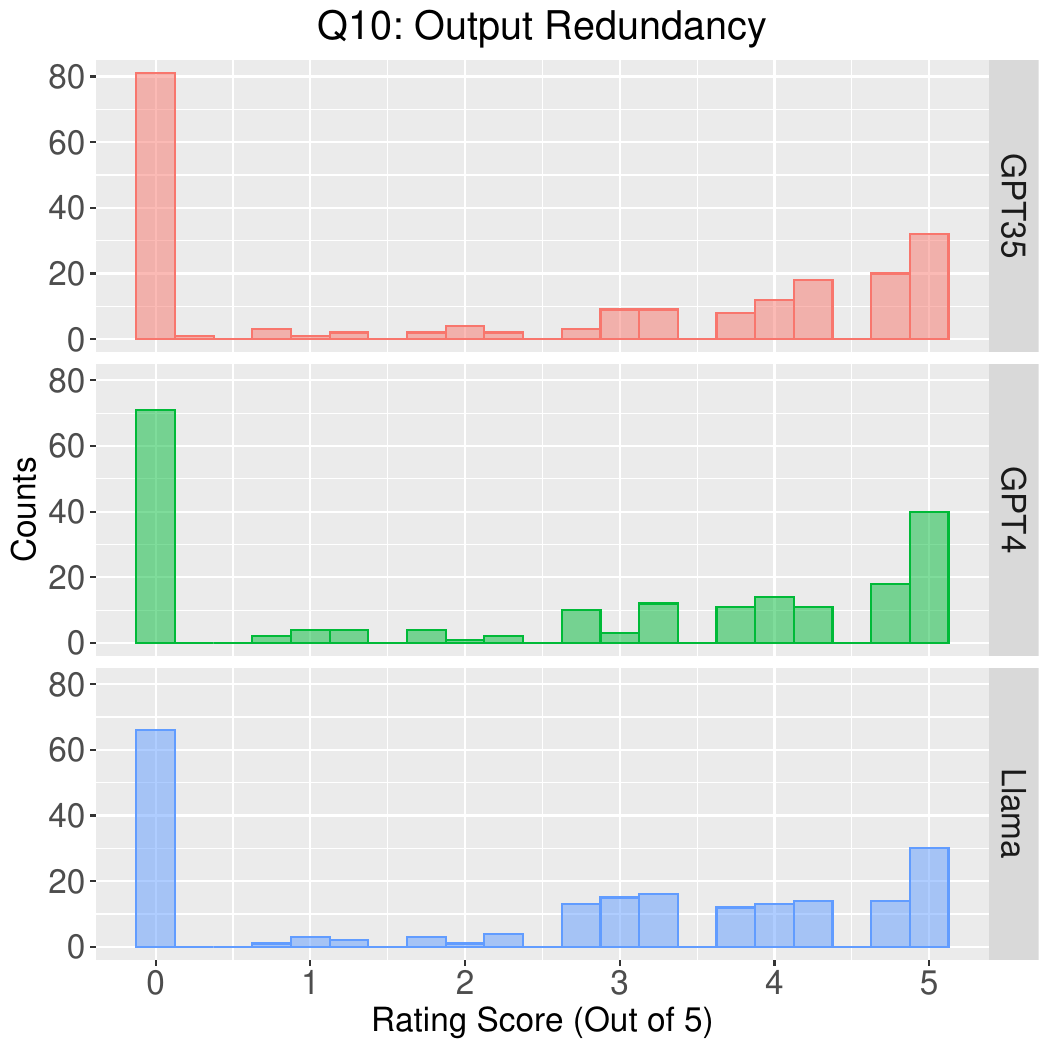} &
			
		\end{tabular}
	\caption{Histograms of individual criterion scores (i.e., $Q_1$ to $Q_{10}$) across three LLMs.}
	\label{fig:ind.hist}
	\end{center}
\end{figure}

This section provides additional details and results. Algorithm~\ref{alg:bootstrap} outlines the steps for the non-parametric bootstrap used to compute CIs for the regression coefficients. Figure~\ref{fig:group.hist} shows the histograms of group scores across the three LLMs. The distribution of Group~1 scores is unimodal but skewed to the left, while Group~2 and Group~3 scores display a bimodal pattern. Figure~\ref{fig:ind.hist} presents the histograms of individual criterion scores across the three LLMs, showing skewed distributions or bimodal behavior in all cases.

\begin{algorithm}
	\caption{Non-parametric bootstrap for computing CIs for regression coefficients.}
	\label{alg:bootstrap}
	\begin{algorithmic}[1]
		\State \textbf{Input:} The original dataset $\D = \{(x_{i}^{M_l}, z_{1il}, z_{2il}): i=1, \dots, 207, l=1, 2, 3\}$,  and the bootstrap sample size $B$.
		\For{$b = 1$ to $B$}
		\State Draw a bootstrap sample $\D_b = \{(x_{i}^{M_l,\ast b}, z_{1il}^{\ast b}, z_{2il}^{\ast b}): i=1, \dots, 207, l=1, 2, 3\}$ with replacement from $\D$.
		\State Fit the regression model with data $\D_b$:
		$$x_{i}^{M_l,\ast b}=\beta_0^{\ast b}+\beta_1^{\ast b}z_{1il}^{\ast b}+\beta_2^{\ast b}z_{2il}^{\ast b}+\veps_{il}^{\ast b}, \quad i=1, \dots, 207, l=1, 2, 3.$$
		\State Store estimated coefficients $(\betahat_0^{\ast b}, \betahat_1^{\ast b}, \betahat_2^{\ast b})$.
		\EndFor
		\State For a regression coefficient, the 95\% bootstrap CI is computed as:
		$$
		\left[ \betahat^{\ast([0.025B])}, \betahat^{\ast([0.975B])} \right],
		$$
where $\{\betahat^{\ast(b)}, b=1,\dots, B\}$ is the sorted version of $\{\betahat^{\ast b}$, $b=1, \dots, B\}$, and $[\,\cdot\,]$ is the rounding function.
		\State \textbf{Output:} Bootstrap CIs for coefficients $\beta_0, \beta_1$, and $\beta_2$.
	\end{algorithmic}
\end{algorithm}

\end{document}